\documentclass[prd, lettersize,twocolumn,superscriptaddress,preprintnumbers,nofootinbib,showpacs]{revtex4-1}
\usepackage{amsmath,amssymb,color,graphicx}
\usepackage{hyperref}

\newcommand{\beq}{\begin{equation}}
\newcommand{\eeq}{\end{equation}}
\newcommand{\be}{\begin{equation}}
\newcommand{\ee}{\end{equation}}
\newcommand{\bea}{\begin{eqnarray}}
\newcommand{\eea}{\end{eqnarray}}
\newcommand{\nn}{\nonumber}

\allowdisplaybreaks[4]

\begin{document}

\title{What can we learn from the total width of the Higgs boson?}

\author{Qing-Hong Cao}
\email{qinghongcao@pku.edu.cn}
\affiliation{Department of Physics and State Key Laboratory of Nuclear Physics and Technology, Peking University, Beijing 100871, China}
\affiliation{Collaborative Innovation Center of Quantum Matter, Beijing 100871, China}
\affiliation{Center for High Energy Physics, Peking University, Beijing 100871, China}

\author{Hao-Lin Li}
\email{lihaolin@itp.ac.cn}
\affiliation{CAS Key Laboratory of Theoretical Physics, Institute of Theoretical Physics, Chinese Academy of Sciences, Beijing 100190, P. R. China}

\author{Ling-Xiao Xu}
\email{lingxiaoxu@pku.edu.cn}
\email{lingxiao.xu@sns.it}
\affiliation{Department of Physics and State Key Laboratory of Nuclear Physics and Technology, Peking University, Beijing 100871, China}
\affiliation{Scuola Normale Superiore, Piazza dei Cavalieri 7, 56126, Pisa, Italy}

\author{Jiang-Hao Yu}
\email{jhyu@itp.ac.cn}
\affiliation{CAS Key Laboratory of Theoretical Physics, Institute of Theoretical Physics, Chinese Academy of Sciences, Beijing 100190, P. R. China}
\affiliation{School of Physical Sciences, University of Chinese Academy of Sciences, No.19A Yuquan Road, Beijing 100049, P.R. China}
\affiliation{Center for High Energy Physics, Peking University, Beijing 100871, China}
\affiliation{School of Fundamental Physics and Mathematical Sciences, Hangzhou Institute for Advanced
Study, UCAS, Hangzhou 310024, China}
\affiliation{International Centre for Theoretical Physics Asia-Pacific, Beijing/Hangzhou, China}

\begin{abstract}
As one of the key properties of the Higgs boson, the Higgs total width is sensitive to global profile of the Higgs boson couplings, and thus new physics would modify the Higgs width. We investigate the total width in various new physics models, including various scalar extension, composite Higgs models, and fraternal twin Higgs model. Typically the Higgs width is smaller than the standard model value due to mixture with other scalar if the Higgs is elementary, or curved Higgs field space for the composite Higgs. On the other hand, except the possible invisible decay mode, the enhanced Yukawa coupling in the two Higgs doublet model or the exotic fermion embeddings in the composite Higgs, could enhance the Higgs width greatly. The precision measurement of the Higgs total width at the high-luminosity LHC can be used to discriminate certain new physics models. 

\end{abstract}

\maketitle

\section{\bf Introduction}

After the discovery of the Higgs boson, the next step is to decipher the particle nature of the Higgs boson, e.g., the mass and width of the Higgs boson ($h$). The Higgs boson mass $m_H$ has been measured very precisely at the Large Hadron Collider (LHC), and its value sheds lights on the stability of the vacuum. On the other hand, the width of the Higgs boson ($\Gamma_h$) is not measured yet. We will focus on in this paper the information we can obtain by precision measurements of the Higgs total width.

At the LHC, information of the total width of the Higgs boson can be extracted from off-shell Higgs production~\cite{Caola:2013yja,Campbell:2013una,Campbell:2013wga,Englert:2014ffa,Logan:2014ppa}, the global fit results of the on-shell Higgs signals~\cite{Barger:2012hv}, $\bar{t}tH$ and four-top productions~\cite{Cao:2016wib}. Measurements of the Higgs total width by the ATLAS and CMS collaboration are found in Refs.~\cite{Aaboud:2018puo,Sirunyan:2019twz}, respectively, assuming standard model (SM)-like Higgs couplings. More interestingly, it is worthwhile noting that the Higgs width can also be measured through the Higgs line shape at $\gamma\gamma$ and $\mu^+\mu^-$ colliders~\cite{Gunion:1992ce,Barger:1996jm}.

Assuming the SM-like Higgs couplings, the total width of the Higgs boson $\Gamma$ can be parametrized as
\begin{align}
\Gamma&\simeq \Gamma_{\text{SM}}\left(c_b^2 \text{Br}_{bb}+c_\tau^2 \text{Br}_{\tau\tau}+c_c^2 \text{Br}_{cc}\right.\nn\\
&\ \ \ \ \ \ +\ c_g^2 \text{Br}_{gg}+c_W^2 \text{Br}_{WW}+c_Z^2 \text{Br}_{ZZ}\nn\\
&\left.\ \ \ \ \ \ +\ c_\gamma^2 \text{Br}_{\gamma\gamma}+c_{\gamma Z}^2 \text{Br}_{\gamma Z}\right)+\Gamma_{\text{invisible}}, 
\end{align}
where the major decay modes of the SM Higgs boson, and the invisible width, are included. The numerical values of the SM total width is $\Gamma_{\text{SM}}\simeq 4.1$ MeV, and the major branching ratios in the SM are $\text{Br}_{bb}\simeq 0.584, \text{Br}_{\tau\tau}\simeq 0.0627, \text{Br}_{cc}\simeq 0.029, \text{Br}_{gg}\simeq 0.0856, \text{Br}_{WW}\simeq 0.214, \text{Br}_{ZZ}\simeq 0.0262$, respectively~\cite{deFlorian:2016spz}. The rescaling factors $c_i$'s denote the effects of new physics (NP) in the Higgs boson decay.
Since $\text{Br}_{\gamma\gamma}$ and $\text{Br}_{\gamma Z}$ are highly suppressed in the SM, we neglect these two decay channels when considering the total width of the Higgs boson. The invisible decay of the Higgs boson takes place in many NP models, therefore, we also include the invisible width $\Gamma_{\text{invisible}}$ in our study. We refer to Ref.~\cite{Brivio:2019myy} for a recent analysis of the Higgs width in the framework of SM effective field theory (EFT).

It remains unknown whether the Higgs boson is fundamental or composite, while the couplings of the Higgs boson to gauge bosons often tend to be smaller than the SM values regardless of this nature. In the case of a fundamental Higgs boson, the decrease of the Higgs-gauge-boson couplings is usually caused by mixing between the Higgs and the new scalar, whereas the exception exist for the Georgi-Machacek model~\cite{Georgi:1985nv,Chanowitz:1985ug} where $c_{W,Z}$ can be enhanced. On the other hand, when the Higgs boson is composite and is the pNGB comes from the global symmetry breaking, the Higgs boson to gauge boson couplings decrease due to the misalignment between the gauged direction and the true vacuum direction where the physical Higgs boson fluctuates around. The above arguments for the decrease of the Higgs boson to gauge boson couplings sometimes also hold for the the Higgs boson to fermion couplings, which leads to a decreased Higgs-boson total width if no invisible Hidden sector decay considered. However, in the two Higgs doublet models (2HDMs)~\cite{Branco:2011iw}, the couplings of the Higgs boson to fermions can be enhanced when deviating from the alignment limit, while in the minimal composite Higgs boson model, an exotic fermion embedding can also leads to the enhanced Yukawa couplings of the top  quark to the Higgs boson. Therefore we investigate several popular models that can modify the Higgs-boson total width significantly.

\begin{figure}[htb!]
\includegraphics[scale=0.28]{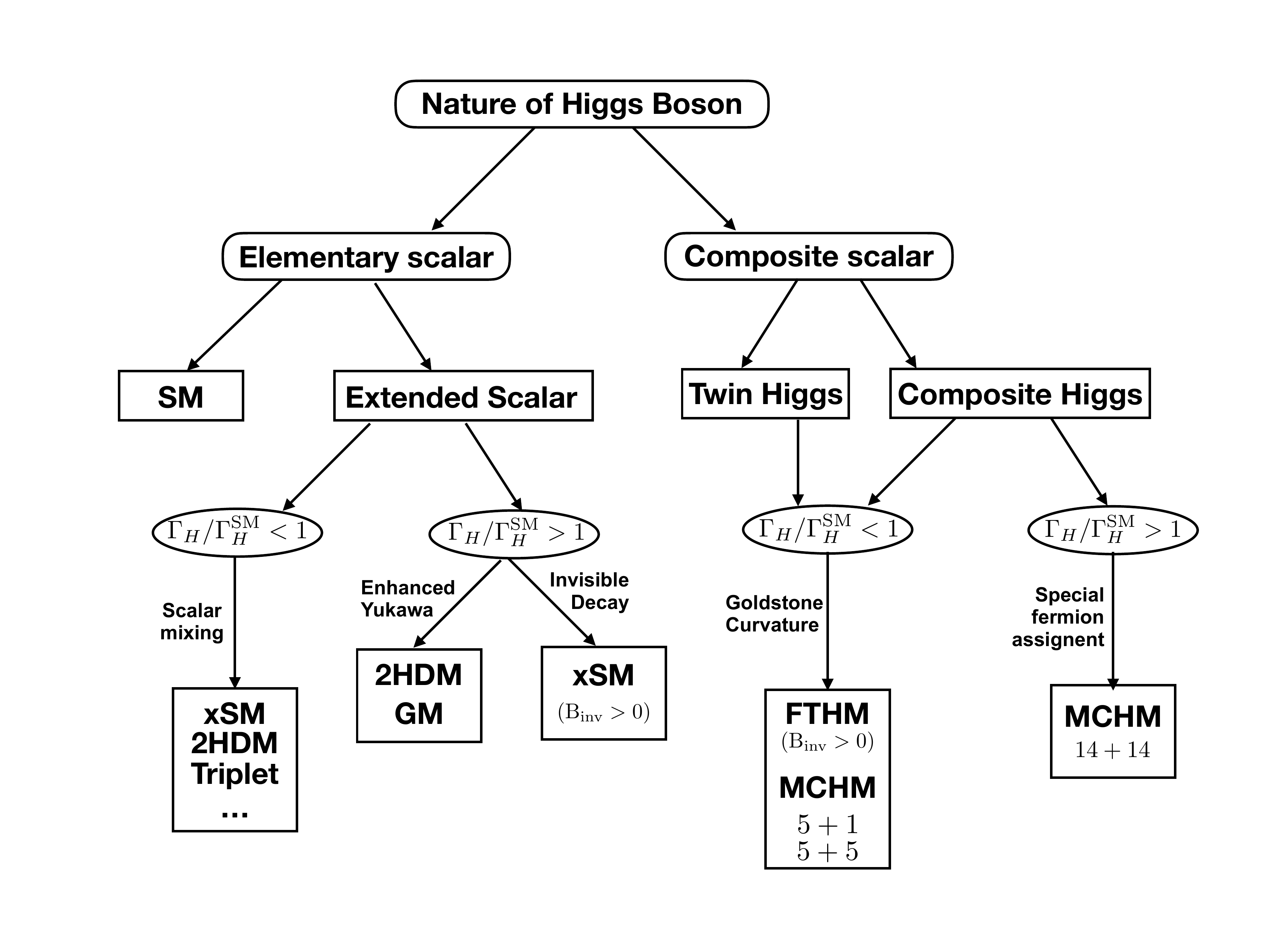}     
\caption{Various kinds of new physics models are classified in the flowchart, based on the nature of the Higgs boson, and future measurement of the total Higgs width.
 }
\label{fig:flow}
\end{figure}

Given the measurement of the total Higgs width, it is possible to discriminate certain classes of new physics models. For the models we studied in this paper, an increased Higgs decay width indicating either an enhanced Yukawa coupling or the existence of invisible decay in the extended scalar models when the Higgs is an elementary particles. When the Higgs is an pNGB, an increased Higgs width usually needs an exotic fermion setup where the SM fermions are embedded in higher dimensional representations.  On the other hand, If one observes a smaller Higgs width compared to the SM, a heavy scalar particle that mixes with Higgs is expected to be found for the Higgs as an elementary particle, such a particle may or may not exist in the pNGB Higgs scenario depending on whether the UV is strong coupled or not.
In Fig.~\ref{fig:flow}, we present that how different kinds of new physics models are classified based on the Higgs width modification and nature of the Higgs boson.

The paper is organized as follows.
In Sec.~\ref{sec:rsm} we discuss the simplest case of the scalar extension, real singlet scalar model~\cite{Profumo:2007wc}. In Sec.~\ref{sec:2hdm} we focus on four types of two Higgs doublet models (2HDMs), which correspond to four different assignments of Yukawa couplings between fermions and Higgs doublets~\cite{Branco:2011iw}. In Sec.~\ref{sec:composite} we focus on minimal composite Higgs boson model~\cite{Agashe:2004rs}, in which the Higgs boson emerges as a pseudo Nambu-Goldstone boson (PNGB) from the coset $SO(5)/SO(4)$. In Sec.~\ref{sec:twinHiggs} we turn our attention to the twin Higgs boson paradigm~\cite{Chacko:2005pe}. In contrast to the original mirror twin Higgs boson model~\cite{Chacko:2005pe}, the fraternal twin Higgs-boson model~\cite{Craig:2015pha} is considered, in which only the third generation of the SM fermions have twin partners. Finally, we conclude.

\section{Real Singlet Model}
\label{sec:rsm}
%%%

The real singlet scalar model is the most simplest extension to the SM scalar sector that is possible to generate strong first order electroweak phase transition. 
It has been extensively studied in the literature~\cite{OConnell:2006rsp,Profumo:2007wc}. In this model, after electroweak symmetry breaking, the real singlet scalar $S$ mixes with the neutral CP-even component $h_0$ in the Higgs boson doublet such that the physical Higgs boson $h$ we observed can be expressed as:
\begin{eqnarray}
h =  h_0\cos\theta + S\sin\theta .
\end{eqnarray}  
Therefore the couplings of the Higgs boson $h$ to other SM particles are scaled by an overall factor $\cos\theta$. If we assume that, in the minimal setup, there is no hidden sector particles that the Higgs boson $h$ can decays to and the mass of the other singlet-like scalar is larger than half of the SM-like Higgs mass, then the Higgs-boson total width is the SM value scaled by $\cos\theta$. In this case, any confirmation of an enhanced Higgs total width can help to ruled out this minimal setup. 
This conclusion could be extended to extended scalar models, such as the general model setup considered in Ref.~\cite{Ramsey-Musolf:2021ldh}.

\begin{figure}
\includegraphics[scale=0.3]{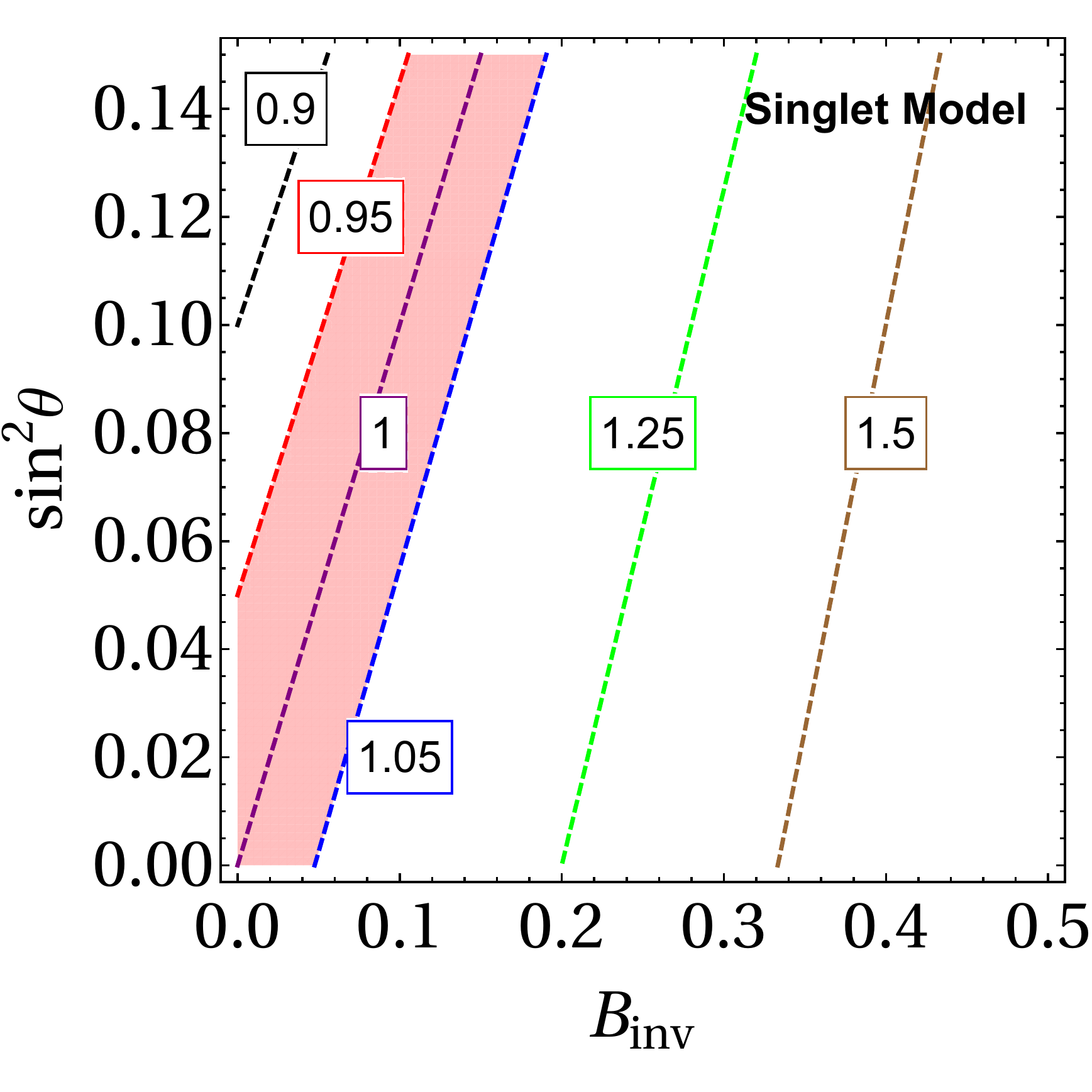}     
\caption{Contour plots on the total width of the Higgs boson in the singlet scalar extended model in the plane of the mixing angle $\sin^2\theta$ and the invisible decay branching ratio $B_{inv}$. The parameter space in the pink region the SM value cannot be differentiated with the SM with the future HL-LHC. 
 }
\label{fig:singlet}
\end{figure}

Figure~\ref{fig:singlet} displays the ratio of the Higgs boson width to its SM value  in the plane of the invisible decay branching ratio $B_{inv}$ and the mixing angle $\sin^2\theta$. The chosen ranges of the two axes are within the current LHC constraints ~\cite{Aad:2019mbh}. The pink region denotes the parameter space where the deviation of the total width is within 5\% of the SM value such that it cannot be discriminated with the SM with the future high luminosity LHC (HL-LHC). However, if the Higgs boson width is more than the SM value by 25\%, then it is likely that there is an invisible decays of the Higgs boson in the singlet model, regardless of the mixing angle $\theta$.

\section{Two Higgs Doublet Models}
\label{sec:2hdm}

We consider the general CP-conserving 2HDM with soft $Z_2$ breaking mass term \cite{Branco:2011iw}. The Higgs potential is
\begin{eqnarray}
 \label{pot}
V(\phi_1, \phi_2)&=&m_{11}^2(\phi_1^\dagger\phi_1)
+m_{22}^2(\phi_2^\dagger\phi_2)-\left(m_{12}^2 (\phi_1^\dagger\phi_2)+{\rm h.c.}\right)
\nonumber \\
&&+ \frac{\lambda_1}{2}(\phi_1^\dagger\phi_1)^2
+\frac{\lambda_2}{2}(\phi_2^\dagger\phi_2)^2+\lambda_3(\phi_1^\dagger\phi_1) (\phi_2^\dagger\phi_2) 
\nonumber \\
&&
+\lambda_4(\phi_1^\dagger\phi_2) (\phi_2^\dagger\phi_1) +\frac{1}{2}\left[\lambda_5(\phi_1^\dagger\phi_2)^2  +{\rm h.c.}\right] \ , 
\label{pot_gen}
\end{eqnarray}
where all the parameters in the above potential are assumed to be real.
Four types of Yukawa interactions can be introduced with different assignments of the $Z_2$ charge of the fermion fields to prevent the tree-level flavor changing neutral current (FCNC). In each model, different types of the right handed fermion fields couple to different Higgs doublets; see Table~\ref{tab:type_yukawa}.

\begin{table}[b]
\caption{Models without tree-level FCNC where the $u_R^i$, $d_R^i$ and $e_R^i$ represents the right-handed up-type quark, down-type quark and charged lepton fields, respectively. In different types of models they couple to different Higgs doublets.}
\label{tab:type_yukawa}
\begin{center}
\begin{tabular}{c|c|c|c||c|c|c|c}  \hline
Model & $u_R^i$     & $d_R^i$& $e_R^i$    & Model & $u_R^i$     & $d_R^i$& $e_R^i$     \\
\hline
Type-I     & $\phi_2$   &$ \phi_2$ & $\phi_2$    & Type-II     & $\phi_2$    &$\phi_1$           & $\phi_1$    \\
\hline
Type-X     &$\phi_2$    & $\phi_2 $         & $\phi_1$  & Type-Y     & $\phi_2$     & $\phi_1$          & $\phi_2$  \\
\hline
\end{tabular}
\end{center}
\end{table}

People usually trade the parameters in the potential with a set of physical parameters: $v$, the electroweak vacuum expectation value (vev); $\tan\beta$, the ratio of the vevs of two Higgs doublets; $\alpha$, the rotation angle which diagonalizes the mass matrix of the CP-even neutral Higgs sector; $m_{1(2,3)}$, the mass for the tree neutral Higgs, where $m_1 =125.1$ GeV is identified with the mass of the SM-like Higgs boson, $m_2$ and $m_3$ are masses of the heavy CP even and CP-odd Higgs boson, respectively; $m_{H^\pm}$, the mass of the charged Higgs boson; $m_{12}^2$, the parameters that is sensitive to several the theoretical bonds of the theory (perturbative unitarity, stability)~\cite{Kling:2016opi}. In addition, we denotes the physical SM-like Higgs boson as $h$ and the charged Higgs boson as $H^\pm$.

The Higgs boson width in the 2HDM is dominantly determined by the parameters $\alpha$, $\beta$, which fix the Higgs couplings to fermions and gauge bosons. The dependence on $m_{12}^2$, $m_2$ and $m_{H^\pm}$ appears when taking into account the partial widths of $h\to\gamma\gamma$ and $h\to Z\gamma$. We define the rescaling of the SM Higgs coupling and the $h$ to $H^\pm$ couplings as follows:
\begin{eqnarray}
{\mathcal L}_{h} &=&
- \sum_{f=u,d,\ell} \frac{m_f}{v} 
c_f {\overline f}f h- \sum_{V=Z,W}\frac{(1+\delta_V)m_V^2}{v}c_V V^\mu V_\mu h \nonumber \\
&&-\frac{g_{c}}{v}hH^+ H^-,  
\end{eqnarray}
where the rescaling factors of the Yukawa couplings $c_f$'s in each models are summarized in Table~\ref{tab:yukawa}, and the $c_V$ and $g_c$ are universal for each model, 
\begin{eqnarray}
c_V &=& \sin(\beta-\alpha), \\
g_c&=&2\left(m_{1}^2-\frac{2m_{12}^2}{\sin 2\beta}\right)\cot 2\beta \cos(\beta-\alpha)\nonumber \\
&+&\left(m_{1}^2+2m_{H^\pm}^2-\frac{4m_{12}^2}{\sin 2\beta}\right)\sin(\beta-\alpha).
\end{eqnarray}

\begin{table}
\caption{The rescaling factor of the Yukawa couplings to the SM value.}
\label{tab:yukawa}
\begin{center}
\begin{tabular}{c|c|c|c}  \hline
 & $c_u $ & $c_d $ & $c_l $     \\
\hline
Type-I    & $c_{\beta-\alpha}/t_\beta+s_{\beta-\alpha}$   & $c_{\beta-\alpha}/t_\beta+s_{\beta-\alpha}$ & $c_{\beta-\alpha}/t_\beta+s_{\beta-\alpha}$      \\
\hline
Type-II & $c_{\beta-\alpha}/t_\beta+s_{\beta-\alpha}$        & $s_{\beta-\alpha}-c_{\beta-\alpha}t_\beta$            & $s_{\beta-\alpha}-c_{\beta-\alpha}t_\beta$   \\
\hline
Type-X & $c_{\beta-\alpha}/t_\beta+s_{\beta-\alpha}$     & $c_{\beta-\alpha}/t_\beta+s_{\beta-\alpha}$          & $s_{\beta-\alpha}-c_{\beta-\alpha}t_\beta$    \\
\hline
Type-Y & $c_{\beta-\alpha}/t_\beta+s_{\beta-\alpha}$     & $s_{\beta-\alpha}-c_{\beta-\alpha}t_\beta$          & $c_{\beta-\alpha}/t_\beta+s_{\beta-\alpha}$   \\
\hline
\end{tabular}
\end{center}
\end{table}

\begin{figure}
\includegraphics[scale=0.24]{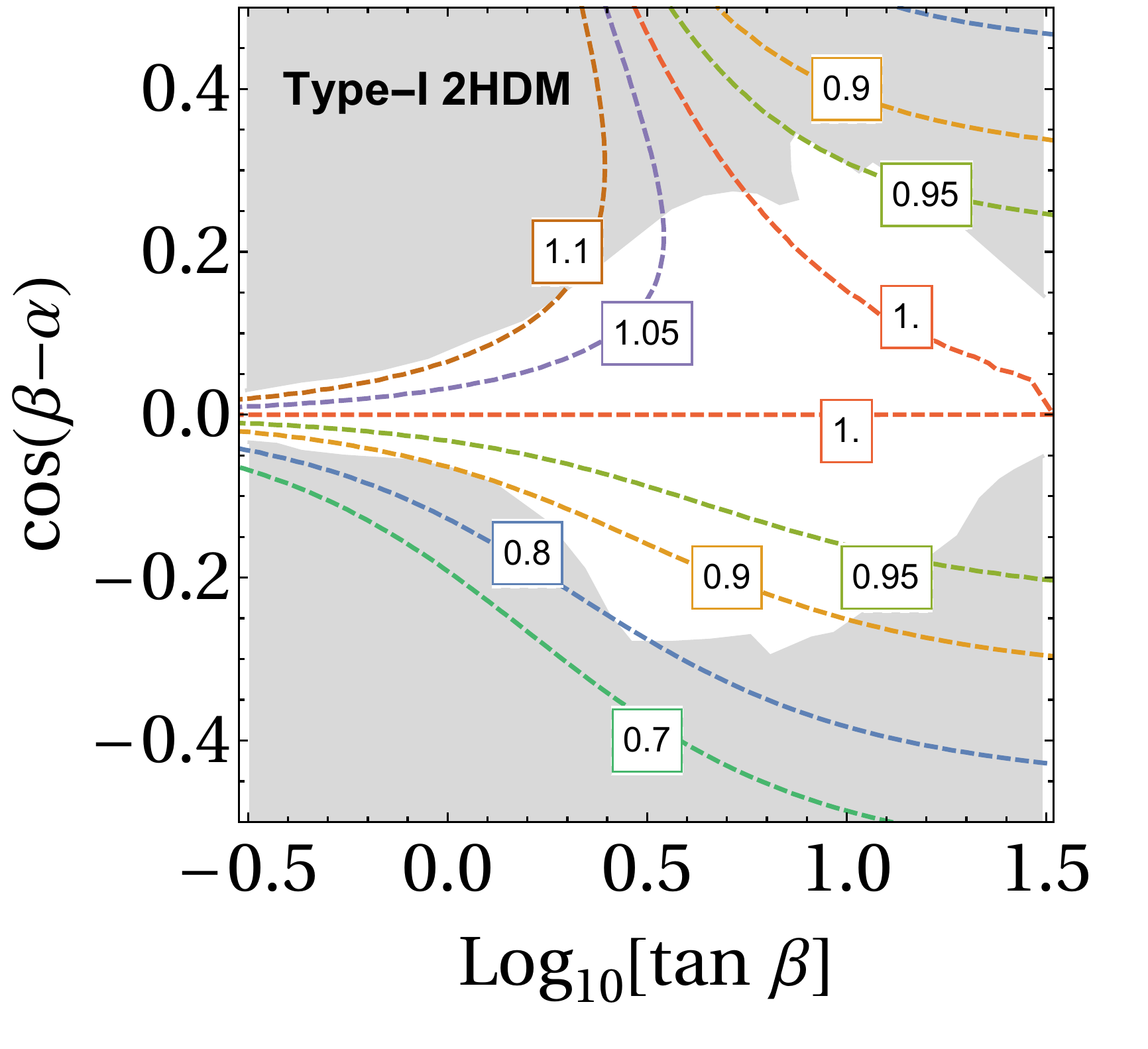}     
\includegraphics[scale=0.24]{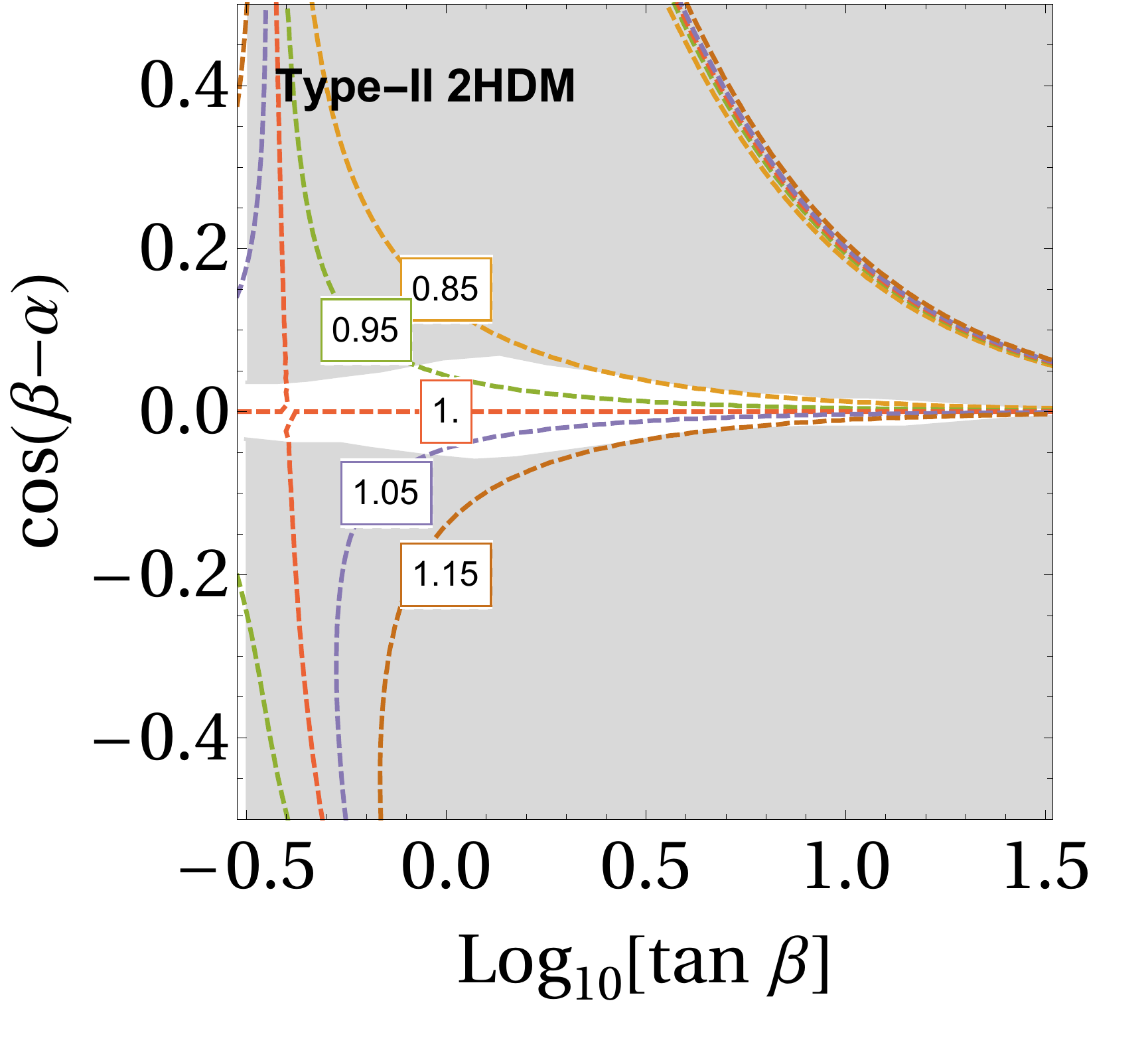}
\includegraphics[scale=0.24]{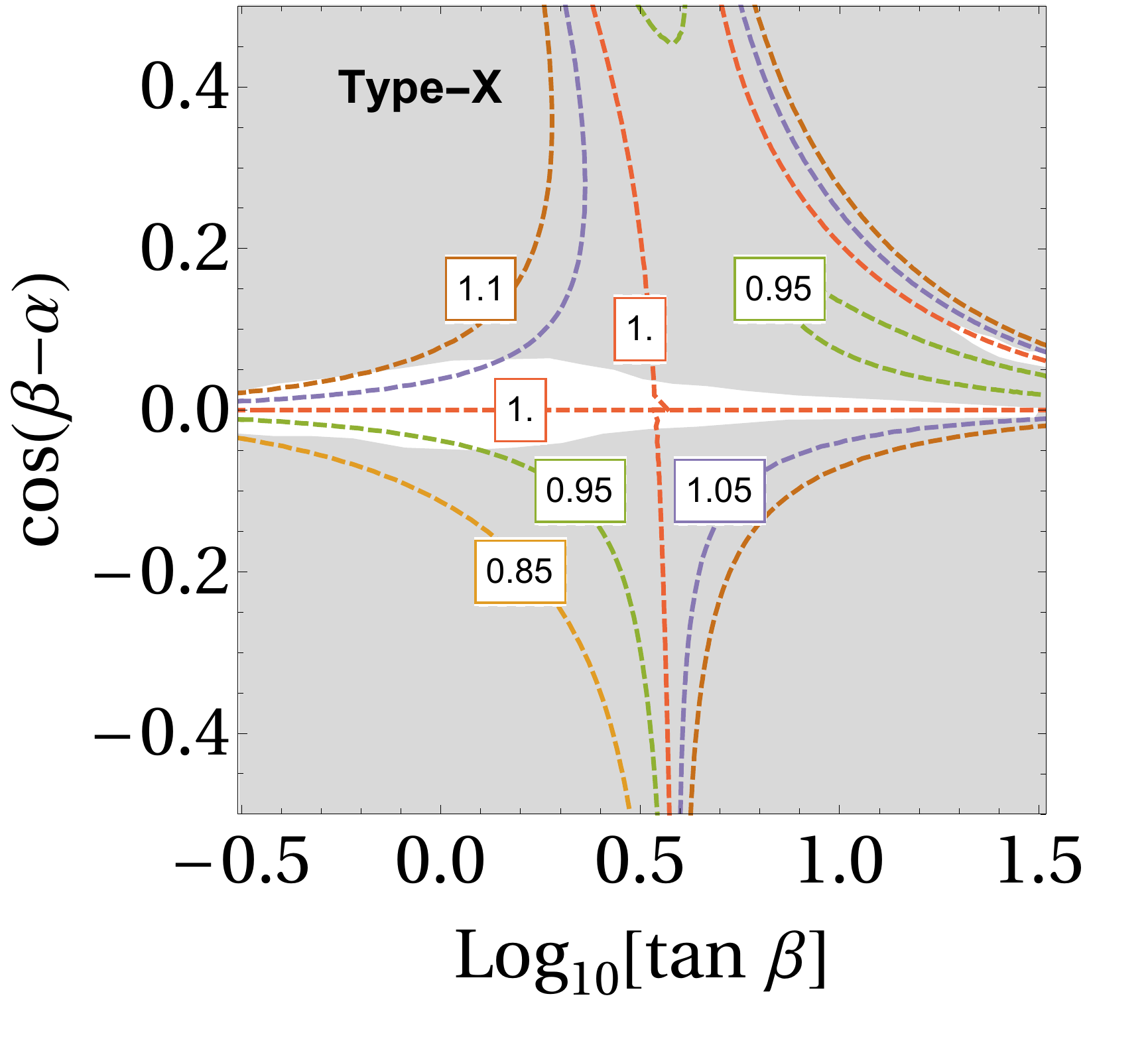}
\includegraphics[scale=0.24]{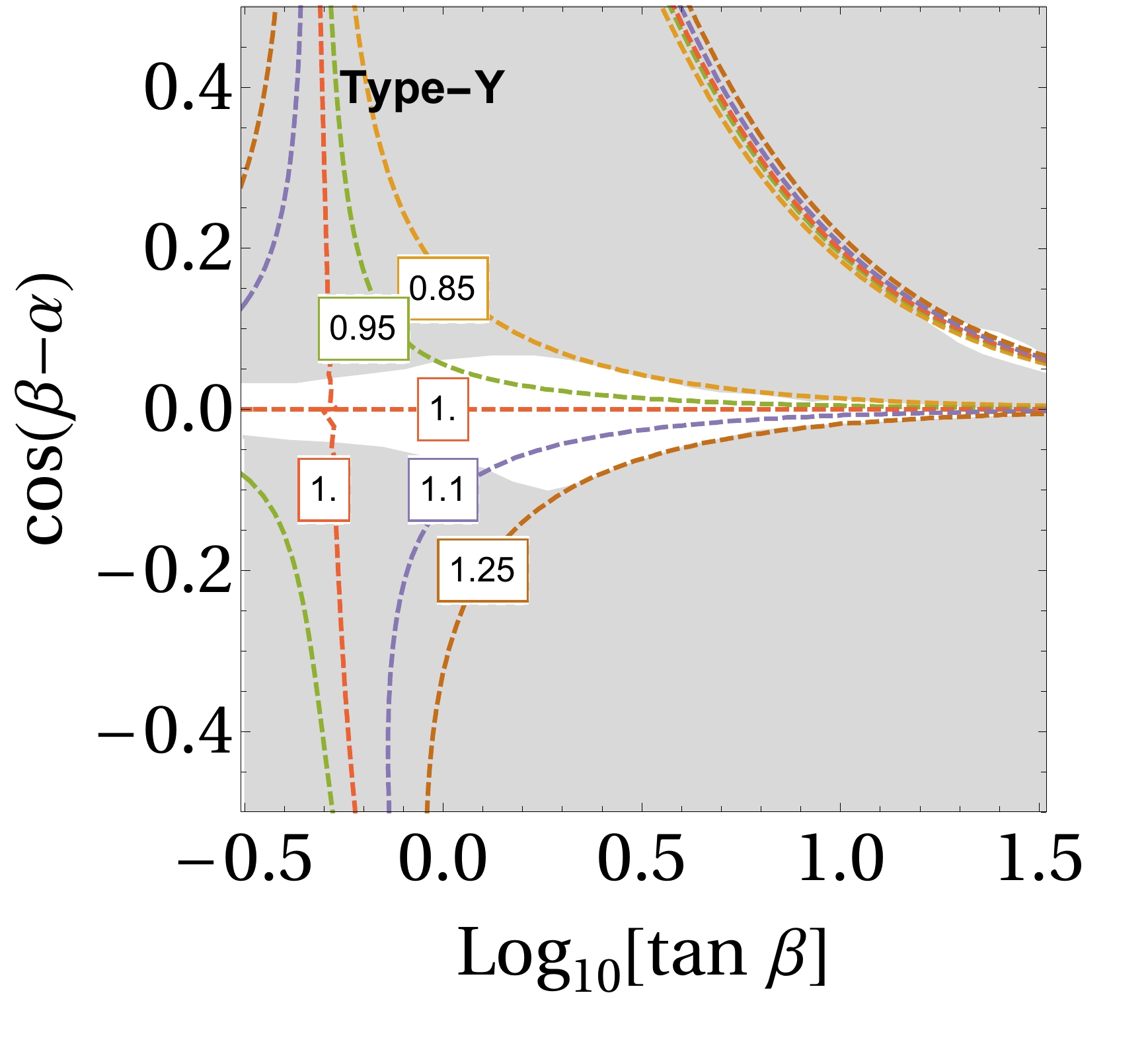}
\caption{Contour of the ratio of the total width of the Higgs boson to the SM value in various 2HDMs in the $\cos(\beta-\alpha)$ vs $\tan\beta$ plane. We have set the mass parameters in each types of model as: $m_{H^\pm}=m_2=m_3=1$ TeV, $m_{12}^2=1$ TeV$^2$. 
 }
\label{fig:2hdmtype}
\end{figure}

We use the public code \texttt{2HDMC-1.7.0}~\cite{Eriksson:2009ws} to calculate the total width of the Higgs boson numerically. Figure~\ref{fig:2hdmtype} plots the contours of the ratio of the total width to the SM value in four types of models in the plane of $\cos(\beta-\alpha)$ and $\tan\beta$. Different color lines represent the contours of the ratio, the gray region is excluded by the global fit of the single Higgs boson production, which is obtained directly from Fig.~1 in Ref.~\cite{Chowdhury:2017aav}. In each model, we have set mass parameters as: $m_{H^\pm}=m_2=m_3=1$ TeV, $m_{12}^2=m_2^2\sin\beta \cos\beta$. The reason that we chose such a $m_{12}^2$ is to satisfy the unitarity constraint in large $\tan \beta$ regions~\cite{Kling:2016opi}.
The mass difference between the charged Higgs boson and the neutral heavy Higgs boson should be less than 300 GeV, i.e., $|m_{H^\pm}-m_{2,3}|\leq 300~{\rm GeV}$~\cite{Chowdhury:2017aav}. The range of the $\tan\beta$ is taken to be (0.3, 32), this region is well consistent with the region allowed by 95.4\% confidential level when taking into account all the available constraints~\cite{Chowdhury:2017aav}.

In general, if one neglects the partial width contributions from $h\to \gamma\gamma$ and $h\to Z\gamma$, then the Higgs-boson total width is uniquely determined by two parameters: $\cos(\beta-\alpha)$ and $\tan\beta$, while the information of other parameters enters via the coupling of the charged Higgs boson to the SM Higgs boson $g_c$ when these two decay channels $h\to \gamma\gamma$ and $h\to Z\gamma$ are taken into account.

Figure~\ref{fig:2hdmtype} shows that ratio of the Higgs-boson total width to the SM value in the 2HDMs roughly ranges from 0.8 to 1.25, and only the Type-I model can have the ratio less than 0.85; on the other hand, only the Type-X model can have the ratio larger than 1.15~. Therefore, all the four types of models are likely to be excluded if the LHC find a ratio larger than 1.25 or less than 0.8~. If one find the ratio is in the range of (1.15,1.25) ((0.8,0.85)), then it must be Type-X (Type-I) model provided the CP conserving 2HDM were realized in the nature.

Figure~\ref{fig:2hdmtype} also demonstrates that in the Type-I model only a positive $\cos(\beta-\alpha)$ combined with a rather small $\tan\beta$  can give an enhanced Higgs total width, the reason is simply that all the fermion couplings are enhanced in this region. However, in the Type-II and Type-X model, an enhanced Higgs total width can only obtained by a negative $\cos(\beta-\alpha)$, the main reason is that the couplings to the bottom quarks are enhanced. In Type-X model, one can obtain an enhanced Higgs total width with either a positive $\cos(\beta-\alpha)$ combined with a relatively small $\tan\beta$  or a negative $\cos(\beta-\alpha)$ combined with a relatively large $\tan\beta$. In the former case, the increase of the Higgs width is mainly due to enhanced couplings to the top quarks which further increases the decay width to gluons, while for the later case, the increase of the Higgs width is mainly caused by the enhanced decay width to bottom quarks.  In addition, in Type-II, Type-X and Type-Y models, one can find a nearly vertical line corresponding to the ratio of the Higgs width equal to one, the reason is that the increase of decay widths of some channels due to increased fermions couplings is exactly compensated by the reduction in decay widths of some other channels caused by decreased Higgs couplings to other fermions and gauge bosons.  

\begin{figure}
\includegraphics[scale=0.23]{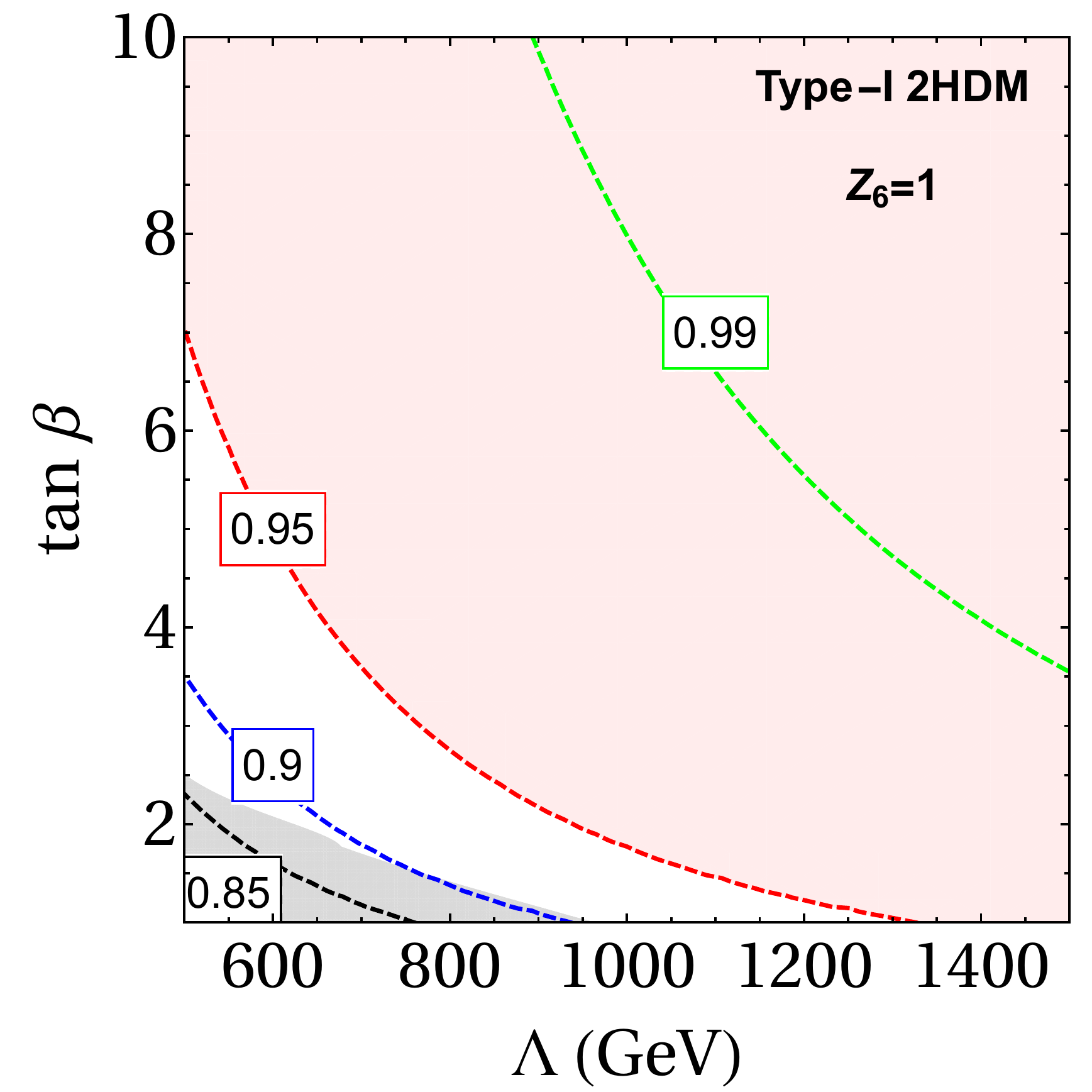}     
\includegraphics[scale=0.23]{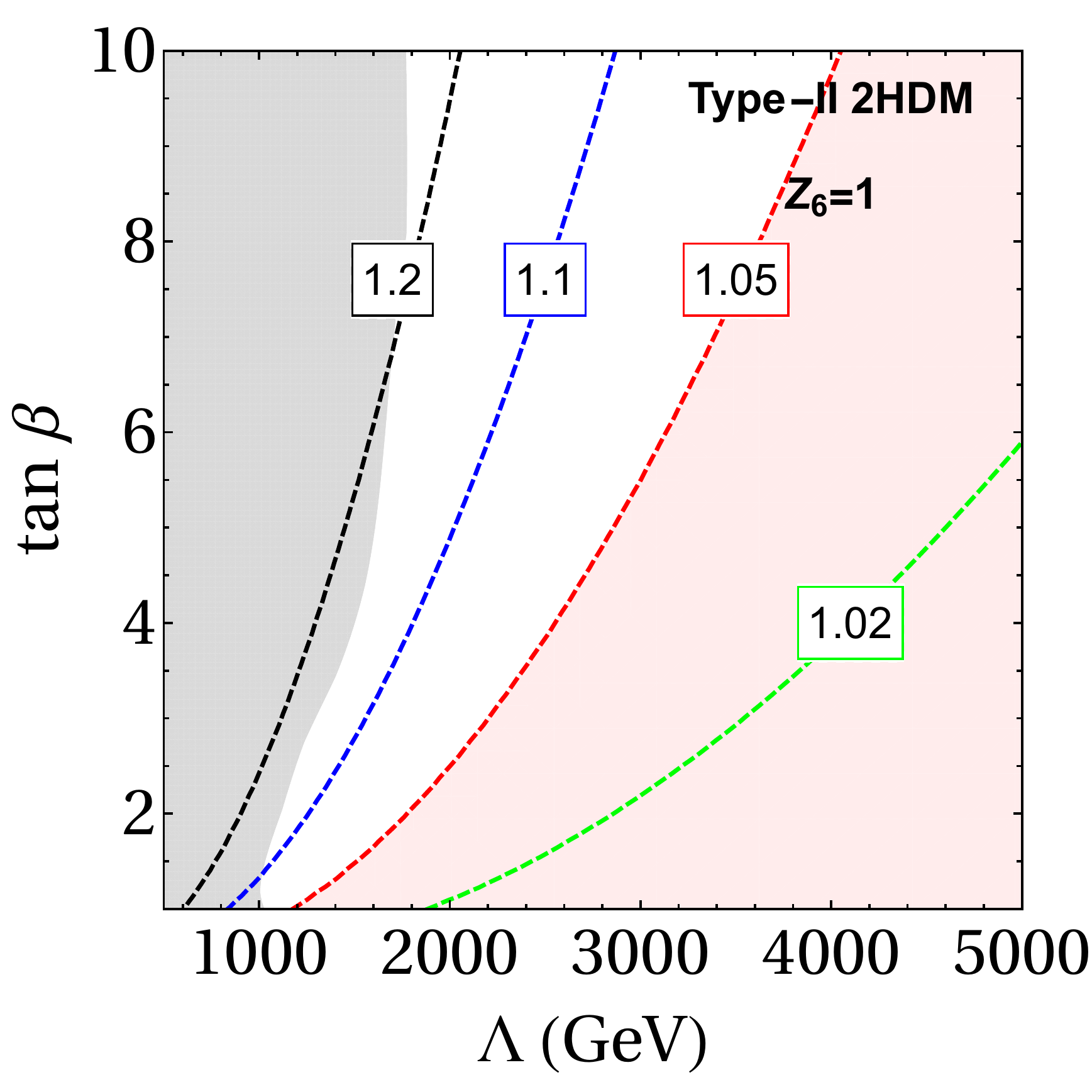}
\includegraphics[scale=0.23]{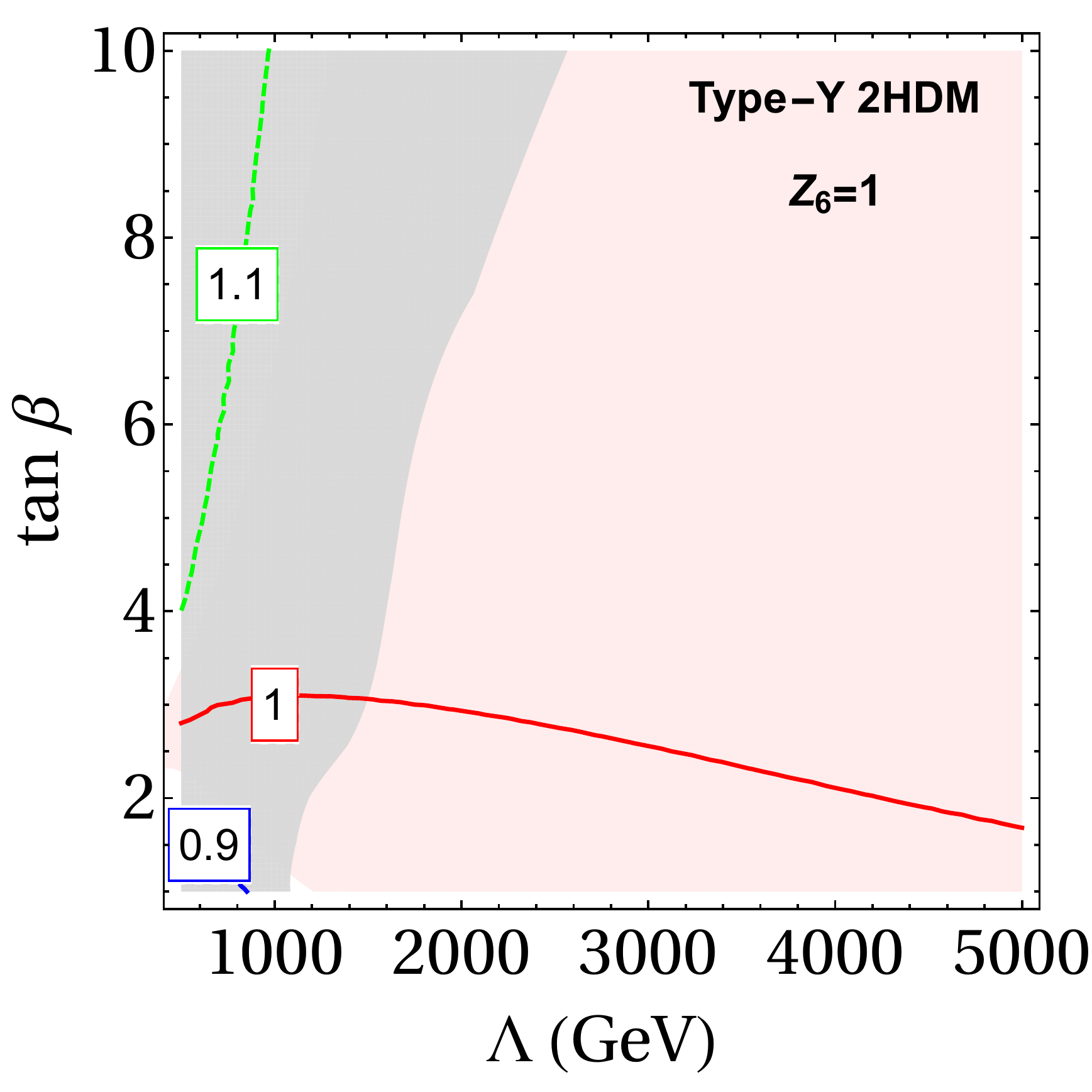}
\includegraphics[scale=0.23]{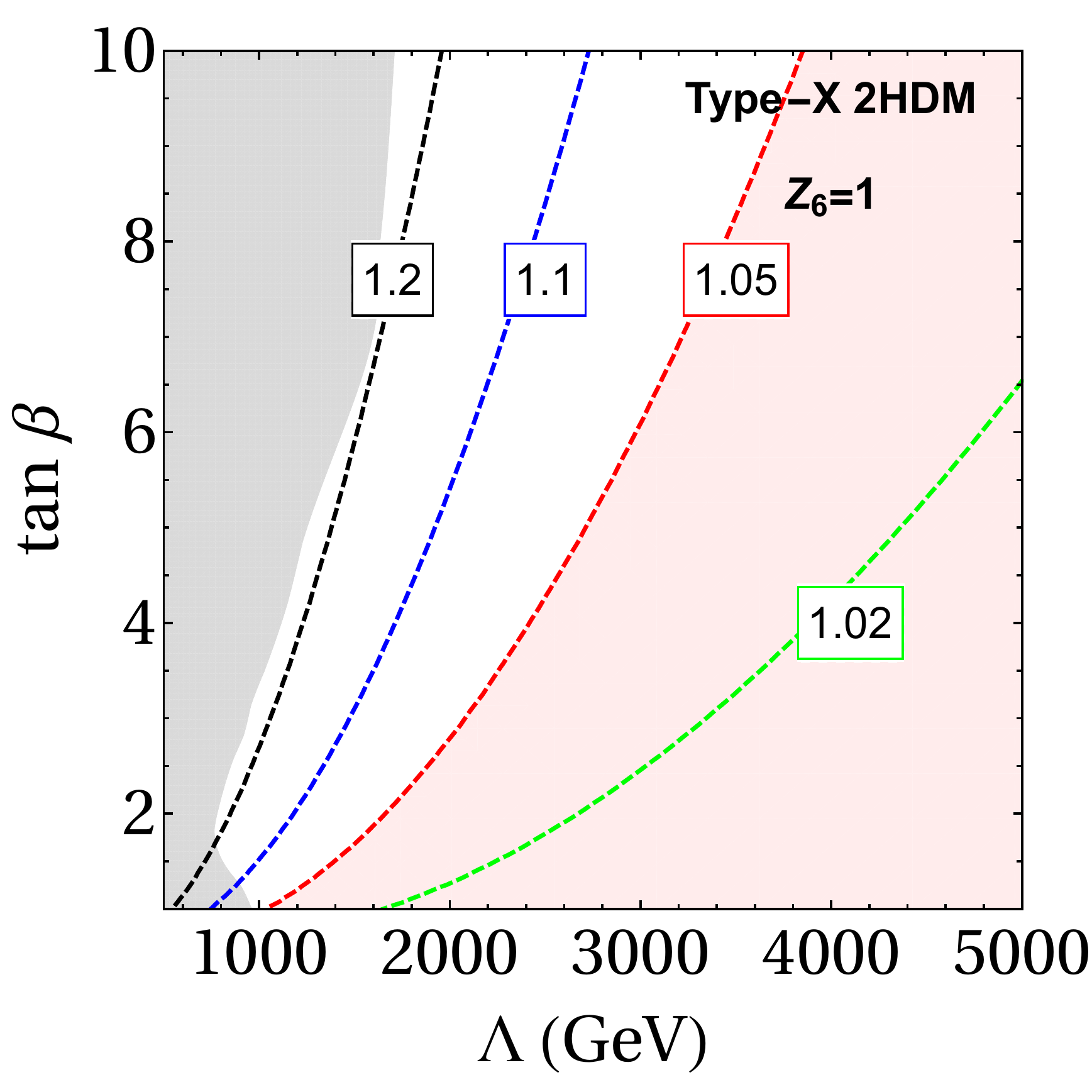}
\caption{Contour of the ratio of the total width of the Higgs boson to the SM value in various 2HDMs  with the parameter $Z_6=1$. The gray regions are excluded by the Higgs global fit, the pink regions are the parameter spaces such that the deviation of the Higgs total width from the SM value are too small to probe at future HL-LHC.
 }
\label{fig:2hdm}
\end{figure}

In the decoupling limit, where we assume the masses of the three Higgs are equal and much larger than the electroweak scale: $m_2=m_3=m_{H^\pm}\sim\Lambda\gg v$. In this case, one can work in an effective field theory derived from the Higgs basis, where $\cos(\beta-\alpha)$ scales as: 
\bea
\cos(\beta-\alpha)=-Z_6\frac{v^2}{\Lambda^2},
\label{eq:cb_a}
\eea
where $Z_6$ is the coefficient of the $|H_1|^2H_1^\dagger H_2$ which can be either positive or negative and $\Lambda^2$ is the coefficient of $|H_2|^2$ in the Higgs basis~\cite{Belusca-Maito:2016dqe}. Then one can translate the bond on the Higgs width onto the new physics scale $\Lambda$ with different value of $\tan\beta$ assuming $Z_6$ a ${\cal O}(1)$ parameter. We plot the contours of the ratio of the Higgs width to the SM value in different models in Fig.~\ref{fig:2hdm} with $Z_6=1$, the gray regions are excluded by the Higgs global fit of the LHC Run1 and Run2 results as above~\cite{Chowdhury:2017aav}, the pink region represents the parameter space in which the deviation of the Higgs total width is within 5\% of the SM value so that the new physics effects cannot be probed by the measurement of the Higgs total width at the HL-LHC.

\section{Georgi-Machacek Model} 

In the Georgi-Machacek Model, the scalar sector of the SM is extended with a real triplet $(\xi^+,\xi^0,\xi^-)$ with hypercharge $Y=0$ and a complex triplet ($\chi^++,\chi^+,\chi^0$) with hypercharge $Y=2$. A good review of the model can be found in Ref~\cite{Hartling:2014xma}. After electroweak symmetry breaking, the Higgs doublet gets a vev $v_H$ and the neutral components of the real and complex triplet get an equal vev $v_\chi$. The relation between the electroweak vev $v$ defined by the Fermi constant $G_F$ and $v_H$ and $v_\chi$ is given by:
\begin{eqnarray}
v^2=v_H^2+8v_\chi^2.
\end{eqnarray}
The rescaling factor for the gauge boson and fermions are given by:
\begin{eqnarray}
c_{W,Z} &=& \frac{e^2}{2s_W^2c_W^2}\left(\cos\alpha \sin\beta-\frac{2\sqrt{6}}{3}\sin\alpha \cos\beta\right)\\
c_f &=& \frac{\cos\alpha}{\sin\beta} ,
\label{eq:cvf}
\end{eqnarray}
where the angle $\alpha$ related to the diagonalization of the mass matrix of the neutral scalars, and $\cos\beta$ is defined by the ratio of $v_H$ and $v$. From Eq.~\ref{eq:cvf}, one can find that the Higgs boson couplings of fermions and gauge bosons can be larger than 1, which may lead to an enhanced Higgs boson width.

\begin{figure}
\includegraphics[scale=.24]{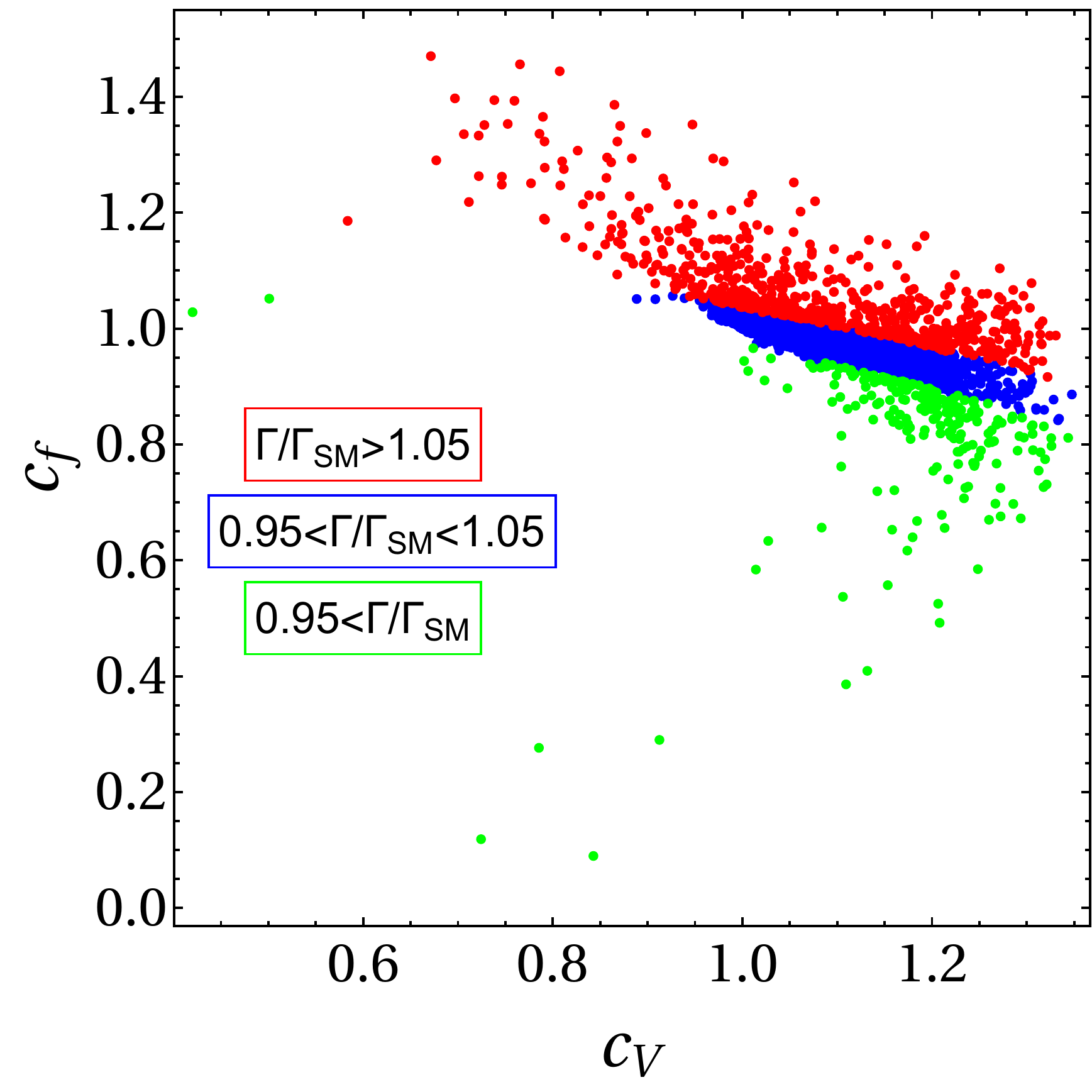}
\includegraphics[scale=.24]{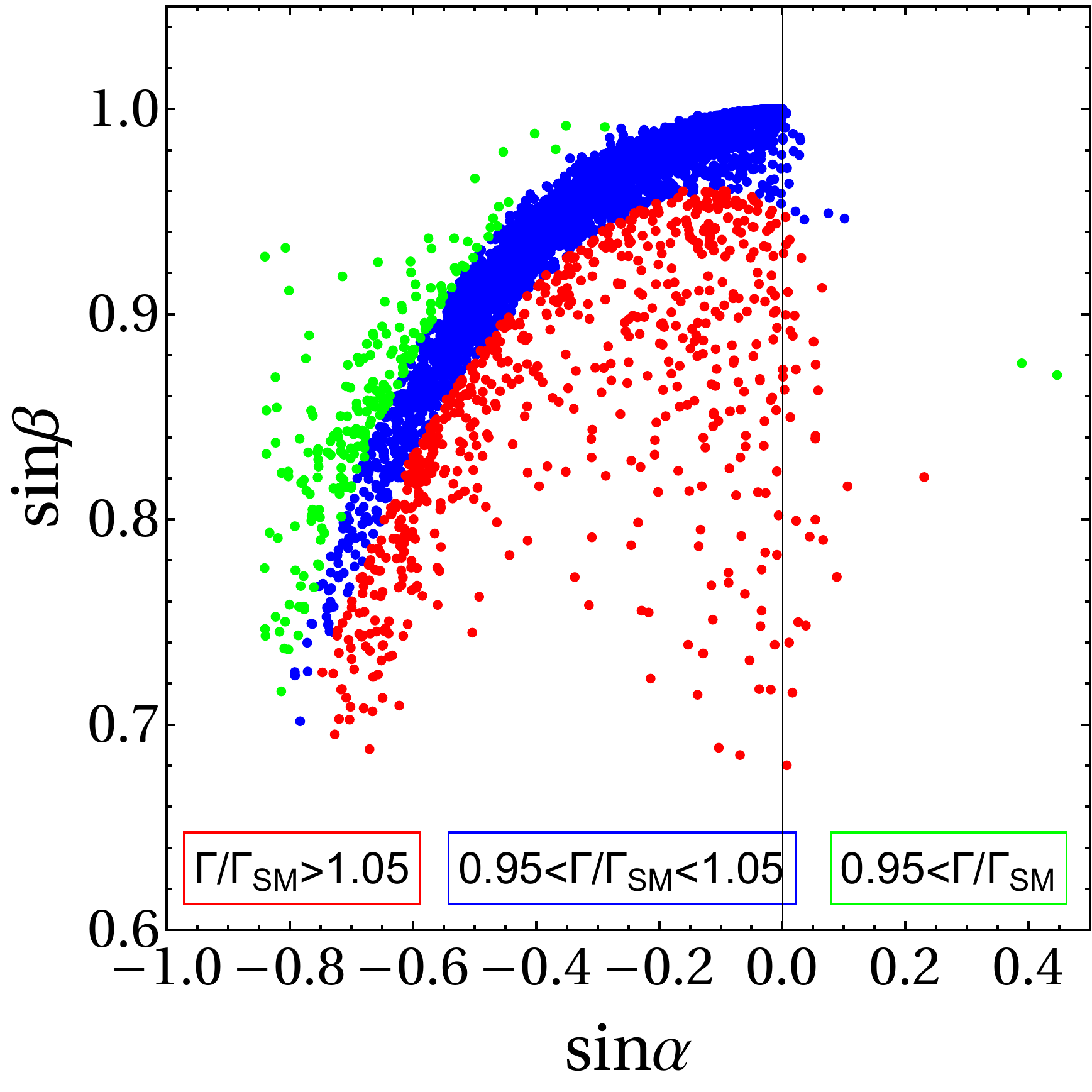}          
\caption{The scattered plot of the scanned points on the $\sin\beta$ vs $\sin\alpha$ plane (left) and $c_f$ vs $c_V$ plane (right), each points satisfies the theoretical bounds and indirect experimental bounds. The blue dots are those with $\Gamma/\Gamma_{SM}$ within [0.95, 1.05], which cannot be distinguished by the future HL-LHC experiment. The red and green dots corresponds to $\Gamma/\Gamma_{SM}>1.05$ and $\Gamma/\Gamma_{SM}<0.95$ respectively and are possible to be probed by the future HL-LHC experiments.}
\label{fig:gmsswd}
\end{figure}

To analyze the Higgs boson width in the model, we use \texttt{GMCALC-1.4.1}~\cite{Hartling:2014xma} to scan the allowed parameter space. The scanning method and the parameter ranges are given in Ref.~\cite{Hartling:2014xma}. Each parameter point in the scan is required to satisfy various constraints: the correct electroweak vev, the stability of the scalar potential, the tree-level unitarity, the experimental bounds on the $S$-parameter and $b\to s\gamma$. Figure~\ref{fig:gmsswd} shows the results of the parameter scan in the plane of $\sin\beta$ and $\sin\alpha$ (left) and in the plane of $c_f$ and $c_V$ (right). The blue dots represents the ratio $\Gamma/\Gamma_{SM}$ within the range $(0.95, 1.05)$ and are not be able to discriminant from the SM value. The red and green dots denotes the ratio $\Gamma/\Gamma_{SM}$ larger than 1.05 and smaller 0.95, respectively, which are possible to be probed by the future experiments. We also find that enhanced fermion couplings strongly correlate with an enlarged Higgs width, while an enlarged Higgs width does not necessarily corresponds to enhanced gauge boson couplings. Especially for a Higgs boson width enlarged by a factor of 1.5 or more, a decreased gauge boson coupling is observed. We also plot the scanned points in the plane of  $\sin\beta$ and $\sin\alpha$. One can find from the plot that these two angle parameters almost determine the value of the Higgs boson width, as the parameter points with different colors are well separated. As one expected, the SM value corresponds to the point with $\sin\beta=1$ and $\cos\alpha=0$, and the red dots corresponding to an increased Higgs boson width tend to have smaller $\sin\beta$'s.

\section{Minimal Composite Higgs Models}
\label{sec:composite}

If the Higgs boson is a pseudo Nambu-Goldstone boson (PNGB) emerging from strong dynamics at TeV scale, its coupling to light fermions $b, c, \tau$ and electroweak gauge bosons $W^\pm, Z$ would be modified with respect to their SM values. In this case, the sign and magnitude of Higgs coupling modifications are mainly dictated by two effects: the Higgs nonlinearity and the composite resonances~\cite{Li:2019ghf}. 

The Higgs nonlinearity denotes the nontrivial curvature of the coset space from which the PNGB Higgs boson emerges~\cite{Alonso:2016btr}. To be concrete, the compact cosets have positive curvatures, while the non-compact ones have negative curvatures. Among various cosets~\cite{Bellazzini:2014yua}, $SO(5)/SO(4)$ receives the most attention, and it is also known as the minimal coset that accommodates the custodial symmetry. Therefore this is called as the minimal composite Higgs model~\cite{Agashe:2004rs}. As a result, the Higgs coupling to the electroweak gauge bosons $W^\pm, Z$ are universally shifted~\cite{Liu:2018vel} as
\bea
c_W=c_Z=\frac{\kappa_{hWW}}{\kappa^{\text{SM}}_{hWW}}=\frac{\kappa_{hZZ}}{\kappa^{\text{SM}}_{hZZ}}=\sqrt{1-\xi}
\eea
at the leading order of the chiral expansion, due to Higgs nonlinearity. Here $\xi\equiv v^2/f^2$ is defined as the ratio of the electroweak scale and the decay constant of the PNGB Higgs boson.

In the fermionic sector, modifications of Higgs boson couplings also depend on fermion embeddings into $SO(5)$ multiplets. In this work, we consider the cases where the top quark and the light fermions $b, c, \tau$ are all embedded in the fundamental representation $5$, or the symmetric tensor representation $14$, of the $SO(5)$ group. Under the paradigm of partial compositeness, there are composite particles mixing with the elementary $t, b, c, \tau$, which are usually considered as the source of explicit $SO(5)$ breaking in the fermionic sector; otherwise the Higgs boson would be a massless exact Goldstone boson when these mixings (as well as gauge couplings) are turned off. After integrating out these composite fermions, they would contribute to the low energy effective couplings of the Higgs boson, through both the chirality-conserving wave functions and the chirality-flipping Yukawa vertices of the elementary fermions. Following Ref.~\cite{Li:2019ghf}, the chirality-conserving wave functions are universally expanded as
\bea
\begin{aligned}
\Pi_{f_L}&=\Pi_{0f_L}+\Pi_{1f_L}\ s^2_h+\Pi_{2f_L}\ s^4_h+\cdots,\ \label{eq:pifL}\\
\Pi_{f_R}&=\Pi_{0f_R}+\Pi_{1f_R}\ s^2_h+\Pi_{2f_R}\ s^4_h+\cdots; \label{eq:pifR}
\end{aligned}
\eea
while the chirality-flipping Yukawa interactions satisfy the expansion
\bea
\Pi_{t_Lt_R}&=\Pi_{1t_Lt_R}\ c_hs_h+\Pi_{2t_Lt_R}\ c_hs^3_h+\cdots\ ,\label{eq:pifLR1}
\eea
or
\bea
\Pi_{t_Lt_R}&=\Pi_{1t_Lt_R}\ s_h+\Pi_{2t_Lt_R}\ s^3_h+\cdots\ ,\label{eq:pifLR2}
\eea
depending on fermion representations. For the cases considered in this work, $(t, b, c, \tau)_{L,R}\subset 5$ and $(t, b, c, \tau)_{L,R}\subset 14$ corresponds to the Eq.~\ref{eq:pifLR1}, while $(t, b, c, \tau)_{L}\subset 5, (t, b, c, \tau)_{R}\subset 1$ corresponds to Eq.~\ref{eq:pifLR2}, as discussed in Ref.~\cite{Li:2019ghf}. 

Accordingly, up to the linear order of $\xi$, the fermion couplings are modified as following:
\begin{align}
c_f&\equiv\frac{\kappa_f}{\kappa_f^{\text{SM}}}\nn\\
&=1-\frac{3}{2}\xi-\xi\left(\frac{\Pi_{1f_L}}{\Pi_{0f_L}}+\frac{\Pi_{1f_R}}{\Pi_{0f_R}}\right)+2\xi \frac{\Pi_{2f_Lf_R}}{\Pi_{1f_Lf_R}}. 
\end{align}
By naive dimensional analysis, 
\bea
\Pi_{1f_L}\sim \frac{y_L^2 f^2}{m_*^2},\ \Pi_{1f_R}\sim \frac{y_R^2 f^2}{m_*^2}, \ \Pi_{0f_L}\sim \Pi_{0f_R}\sim 1,
\eea
where the mass scale $m_*\sim g_* f$ and $y_{L,R}$ are the mixing parameters, while 
\bea
\Pi_{1f_Lf_R}\sim \frac{y_L y_R f^2}{m_*}, \ \Pi_{2f_Lf_R}\sim \frac{y_L y_R f^2}{m_*}.
\eea
For the light fermions $b, c, \tau$, due to their small masses $m_{b, c, \tau}\sim \frac{y_L y_R}{g_*} f$, the mixing parameters $y_{L,R}$ are expected to be much smaller than the coupling strength $g_*$ if the decay constant $f$ is at the order of $1$ TeV. As a result, one can neglect the effect of composite resonances in the wave functions of $b, c, \tau$. Furthermore, $\Pi_{2f_Lf_R}$ does not vanish only in the case of $(t, b, c, \tau)_{L,R}\subset 14$. 

The effective coupling between the Higgs boson and the gluon is, up to the linear order of $\xi$, 
\begin{align}
c_g&\equiv \frac{\kappa_{hgg}}{\kappa^{\text{SM}}_{hgg}}\nn\\
&=1-\frac{3}{2}\xi+2\xi \frac{\Pi_{2t_Lt_R}}{\Pi_{1t_Lt_R}}+\sum_{f=b, c, \tau}\xi\left(\frac{\Pi_{1f_L}}{\Pi_{0f_L}}+\frac{\Pi_{1f_R}}{\Pi_{0f_R}}\right)\ .
\end{align}
Note that contribution of the composite resonances in the $b, c, \tau$ sectors (but not the light fermions $b, c, \tau$) are included in the above equation~\cite{Liu:2017dsz}.

\begin{figure}
\includegraphics[scale=0.35]{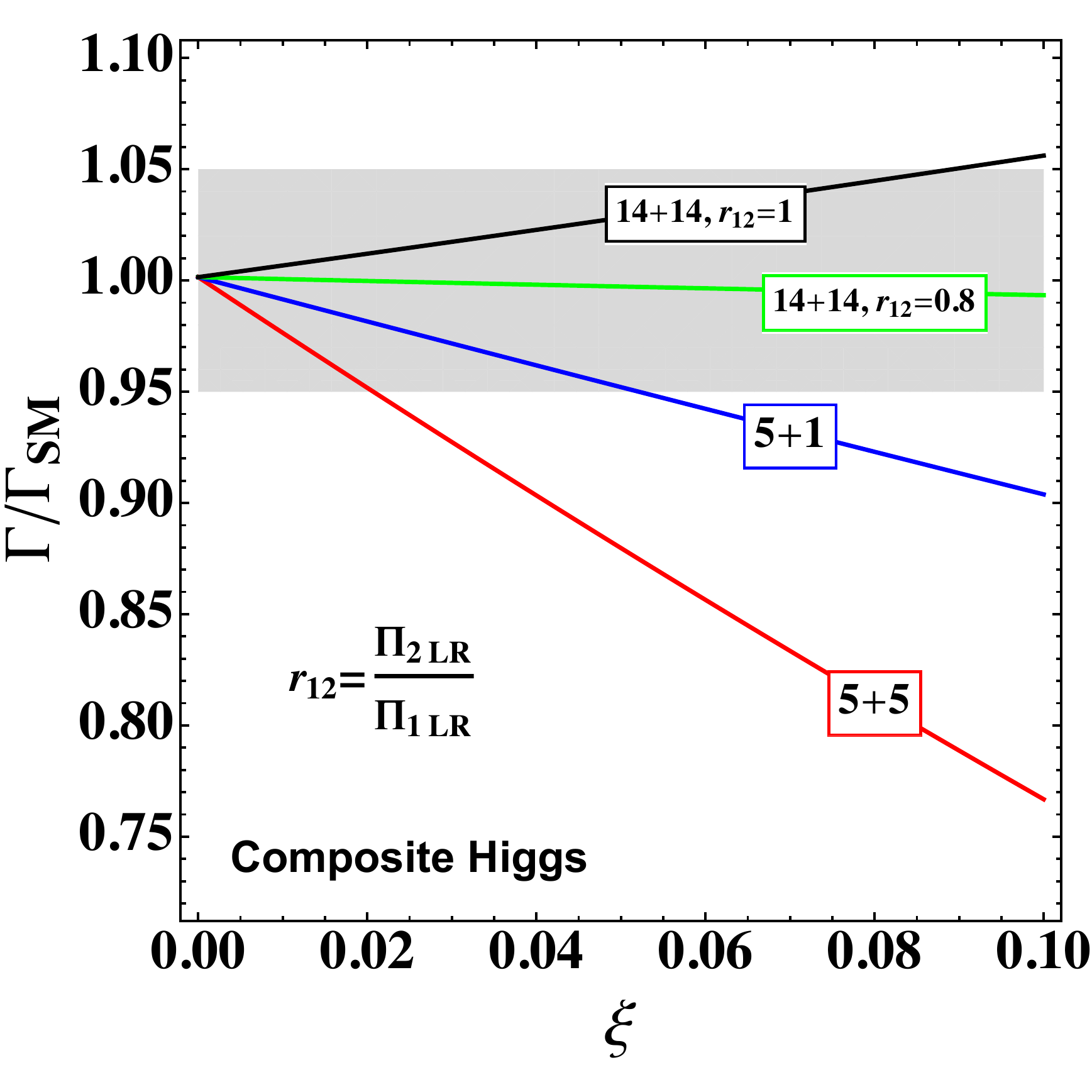}     
\caption{The total width of the Higgs boson in minimal composite Higgs models, in which the left-handed and right-handed fermions are embedded in different representations of the $SO(5)$ group. 
 }
\label{fig:comp}
\end{figure}

Therefore, the Higgs effective couplings in the fermionic sectors are concretely
\begin{itemize}
\item $(t, b, c, \tau)_{L,R}\subset 5$: 
\bea
c_{b, c, \tau}=c_g\simeq 1-\frac{3}{2}\xi\ .
\eea
\item $(t, b, c, \tau)_{L}\subset 5$ and $(t, b, c, \tau)_{R}\subset 1$: 
\bea
c_{b, c, \tau}=c_g\simeq 1-\frac{1}{2}\xi\ .
\eea
\item $(t, b, c, \tau)_{L,R}\subset 14$: 
\bea
c_{b, c, \tau}=c_g\simeq 1-\frac{3}{2}\xi+2\ r_{12}\ \xi\ ,
\eea
where $r_{12}$ is the ratio of $\Pi_{2f_Lf_R}$ to $\Pi_{1f_Lf_R}$. 
\end{itemize}
With the above results, we obtain the total width of the Higgs boson in minimal composite Higgs models, as shown in Fig.~\ref{fig:comp}.
We see explicitly that $\xi\to 0$ corresponds to the decoupling limit where one cannot distinguish a composite Higgs boson from an elementary one. For small fermion representation, the total width of the Higgs boson tends to be smaller than the value in the SM; while for symmetric tensor representation, the total width could be larger.

\section{Fraternal Twin Higgs Model}
\label{sec:twinHiggs}

It is fascinating to consider models in which the Higgs boson could decay into invisible particles. Typically invisible decay exist in neutral naturalness models~\cite{Chacko:2005pe,Craig:2014aea,Xu:2018ofw}. One of these typical model is the twin Higgs scenario~\cite{Chacko:2005pe}, where the naturalness problem is addressed with some particles which are not charged under the SM gauge groups.

In this section we consider the fraternal twin Higgs model~\cite{Craig:2015pha}. In contrast to the original mirror twin Higgs setup~\cite{Chacko:2005pe}, the fraternal model is more minimal and cosmologically safe. The model ingredients of the twin sector consists of one additional Higgs doublet, twin fermions (including $\widetilde{t}_R, \widetilde{b}_R, \widetilde{\tau}_R, \widetilde{Q}_L, \widetilde{L}_L$), twin weak gauge bosons from the $SU(2)$ gauge group in the twin sector, and twin gluons $\widetilde{g}$~\cite{Craig:2015pha}. The Higgs boson is identified as a pseudo Nambu-Goldstone boson from the coset $SU(4)/SU(3)$, or equivalently $SO(8)/SO(7)$, depending whether the custodial symmetry is included in the unbroken group. After the global symmetry breaking, three of the Goldstone bosons would be eaten by the twin weak gauge bosons, which makes their masses roughly at the order of symmetry-breaking scale $f$. For reasonable values of $f$, these twin gauge bosons can be sufficiently heavy.
Only the other light degrees of freedom, lighter than $m_h/2$, could contribute to the invisible width of the Higgs boson. On top of this, all the SM couplings are modified due to the Higgs nonlinearity.

Up to the linear order of $\xi=v^2/f^2$, all the relevant Higgs couplings are as follows: i) for the couplings in the SM sector, 
\begin{align}
c_{b, c, \tau}&=c_g\simeq 1-\frac{1}{2}\xi\ , \nn\\
c_W&=c_Z=\sqrt{1-\xi};
\end{align}
ii) for the couplings in the twin sector, 
\bea
c_{\widetilde{b}, \widetilde{\tau}}=c_{\widetilde{g}} \equiv \frac{\kappa_{h\widetilde{b}\widetilde{b}, h\widetilde{\tau}\widetilde{\tau}, h\widetilde{g}\widetilde{g}}}{\kappa^{\text{SM}}_{hbb, h\tau\tau, hgg}} \simeq \sqrt{\xi},
\eea
assuming the Yukawa couplings are the same, i.e. $y_{t,b,\tau}=y_{\widetilde{t},\widetilde{b},\widetilde{\tau}}$.
All the above Higgs boson  couplings in the twin sector are correlated to their counterparts in the SM sector, in order to realize the naturalness condition~\cite{Li:2019ghf}; as one can see that $c^2_{\widetilde{b}, \widetilde{\tau},\widetilde{g}}+c^2_{b, \tau, g}$ does not depend on $\xi$. 
Furthermore, the masses of the twin fermions are also connected to the partners in the SM sector, e.g.,  
\bea
\frac{m_{\widetilde{t}, \widetilde{b}, \widetilde{\tau}}}{m_{t,b,\tau}} \simeq \frac{\sqrt{1-\xi}}{\sqrt{\xi}}\sim \frac{f}{v}, 
\label{eq:fhw1}
\eea
assuming the Yukawa couplings are the same in the SM and the twin sectors. The twin gluons are massless. They could form glueballs by themself or other bound states with twin fermions below the scale of twin confinement~\cite{Craig:2015pha}.
Note that $m_{\widetilde{b}}$ and $m_{\widetilde{\tau}}$ can be larger or smaller than $m_h/2$ depending on the scale $f$, which would result in different invisible widths of the Higgs boson.

The particles in the twin sector induce the invisible width as following:
\begin{align}
\Gamma_{\text{invisible}}\simeq \Gamma_{\text{SM}} \xi &\Bigg[~~\text{Br}_{gg}\nn\\
&+ \text{Br}_{\tau\tau} \left(\frac{1-x_{\widetilde{\tau}}}{1-x_\tau}\right)^{\frac{3}{2}} \Theta(m_h-2m_{\widetilde{\tau}})\nn\\
&+ \text{Br}_{bb}\left(\frac{1-x_{\widetilde{b}}}{1-x_b}\right)^{\frac{3}{2}}\Theta(m_h-2m_{\widetilde{b}})\Bigg],
\label{eq:fhw2}
\end{align}
where $x_i\equiv 4m_i^2/m_h^2$ for the $i$-particle and $\text{Br}_{jj}$ denotes the branching ratio of the decay channel of $h\to jj$ in the SM. 

\begin{figure}[h!]
\includegraphics[scale=0.3]{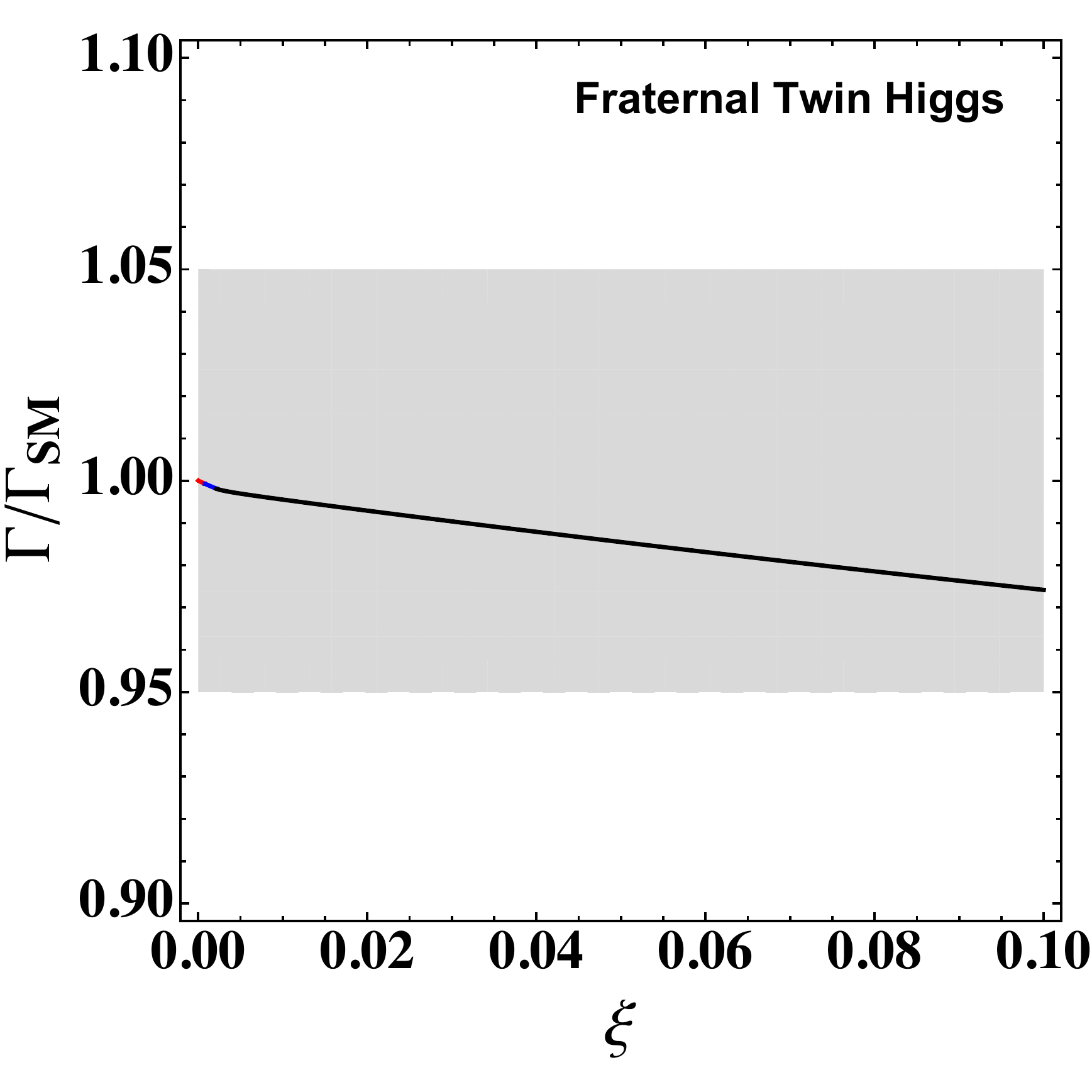}     
\caption{The total width of the Higgs boson in fraternal twin Higgs model. }
\label{fig:twin}
\end{figure}

Note that the running masses of the fermions $b$ and $\tau$ in the SM model are roughly $m_b(M_Z)\simeq 2.9$ GeV and $m_\tau (M_Z)\simeq1.746$ GeV, respectively. We calculate the total width of the Higgs boson in the fraternal twin Higgs model from Eq.~\ref{eq:fhw1} and Eq.~\ref{eq:fhw2}.  Figure~\ref{fig:twin} plots the ratio of the total width of the Higgs boson to the SM value as a function of $\xi$. We notice that the total width is moderately smaller than the SM value, therefore, it is very hard to distinguish the fraternal twin Higgs model from the SM with only the information of Higgs total width.

\section{Conclusion}

The nature of the Higgs boson remains as one unsolved puzzle nowadays. Motivated from the future precise measurement of the Higgs width in the future HL-LHC, we investigated that if the Higgs is elementary or composite, how the model setup could affect the Higgs total width. Typically the Higgs total width is expected to be smaller than the standard model values because of the mixture with other scalars or curved Higgs field space for the composite Higgs. If there is invisible decay channels, the Higgs width could be enhanced, as expected. We further showed that there are also cases that the Higgs width is enhanced due to model setup. 

Depending on the model setup, the following results are in order: 
\begin{itemize}
\item For the singlet extended model, and typical scalar extension models other than the doublet and GM models, the Higgs width is smaller than the SM if no other invisible decay presents. 
\item In the general two Higgs doublet models, the Higgs width can be larger or smaller than the SM value, depending on values of the model parameters. It depends on the concrete value of the Higgs width measured to determine whether a given type of the two Higgs doublet models are excluded or not. 
\item In the Georgi-Machacek model, one will find that an increased Higgs boson coupling always accompanies an enhance Higgs to fermion couplings.
\item For the case of a pseudo-Nambu-Goldstone Higgs boson, including both the minimal composite Higgs model and the twin Higgs model, the Higgs total width is strongly preferred to be smaller than the SM prediction except for the case where the left handed and right handed top quark are both embedded in the 14 representation of SO(5). Therefore, only if the Higgs width is measured to be smaller than the SM value, the minimal composite Higgs model and the twin Higgs model are favored.
\end{itemize}
Overall, if the Higgs total width is measured to be significantly larger or smaller than the SM value, it is possible for us to falsify several new physics models considered above, although the caveat still exists, in which new physics might contain some new hidden sector we never consider and thus the SM Higgs can decay invisibly.

We note that the Higgs total width is sensitive to almost all kinds of Higgs couplings. On the other hand, the Higgs production and partial decay width provide different kinds of information on the Higgs coupling. We expect that combining the Higgs coupling information and the Higgs total width information at the HL-LHC, the Higgs sector could be better understood in near future.

\begin{acknowledgements}
The work is supported in part by the National Science Foundation of China under Grant Nos. 11725520, 11675002, 11635001, 12022514, 11875003 and 12047503. 
JHY is also supported by the National Key Research and Development Program of China under Grant No. 2020YFC2201501. 
\end{acknowledgements}

\bibliographystyle{apsrev}
\bibliography{reference}

\begin{thebibliography}{35}
\expandafter\ifx\csname natexlab\endcsname\relax\def\natexlab#1{#1}\fi
\expandafter\ifx\csname bibnamefont\endcsname\relax
  \def\bibnamefont#1{#1}\fi
\expandafter\ifx\csname bibfnamefont\endcsname\relax
  \def\bibfnamefont#1{#1}\fi
\expandafter\ifx\csname citenamefont\endcsname\relax
  \def\citenamefont#1{#1}\fi
\expandafter\ifx\csname url\endcsname\relax
  \def\url#1{\texttt{#1}}\fi
\expandafter\ifx\csname urlprefix\endcsname\relax\def\urlprefix{URL }\fi
\providecommand{\bibinfo}[2]{#2}
\providecommand{\eprint}[2][]{\url{#2}}

\bibitem[{\citenamefont{Caola and Melnikov}(2013)}]{Caola:2013yja}
\bibinfo{author}{\bibfnamefont{F.}~\bibnamefont{Caola}} \bibnamefont{and}
  \bibinfo{author}{\bibfnamefont{K.}~\bibnamefont{Melnikov}},
  \bibinfo{journal}{Phys. Rev.} \textbf{\bibinfo{volume}{D88}},
  \bibinfo{pages}{054024} (\bibinfo{year}{2013}), \eprint{1307.4935}.

\bibitem[{\citenamefont{Campbell
  et~al.}(2014{\natexlab{a}})\citenamefont{Campbell, Ellis, and
  Williams}}]{Campbell:2013una}
\bibinfo{author}{\bibfnamefont{J.~M.} \bibnamefont{Campbell}},
  \bibinfo{author}{\bibfnamefont{R.~K.} \bibnamefont{Ellis}}, \bibnamefont{and}
  \bibinfo{author}{\bibfnamefont{C.}~\bibnamefont{Williams}},
  \bibinfo{journal}{JHEP} \textbf{\bibinfo{volume}{04}}, \bibinfo{pages}{060}
  (\bibinfo{year}{2014}{\natexlab{a}}), \eprint{1311.3589}.

\bibitem[{\citenamefont{Campbell
  et~al.}(2014{\natexlab{b}})\citenamefont{Campbell, Ellis, and
  Williams}}]{Campbell:2013wga}
\bibinfo{author}{\bibfnamefont{J.~M.} \bibnamefont{Campbell}},
  \bibinfo{author}{\bibfnamefont{R.~K.} \bibnamefont{Ellis}}, \bibnamefont{and}
  \bibinfo{author}{\bibfnamefont{C.}~\bibnamefont{Williams}},
  \bibinfo{journal}{Phys. Rev.} \textbf{\bibinfo{volume}{D89}},
  \bibinfo{pages}{053011} (\bibinfo{year}{2014}{\natexlab{b}}),
  \eprint{1312.1628}.

\bibitem[{\citenamefont{Englert et~al.}(2015)\citenamefont{Englert, Soreq, and
  Spannowsky}}]{Englert:2014ffa}
\bibinfo{author}{\bibfnamefont{C.}~\bibnamefont{Englert}},
  \bibinfo{author}{\bibfnamefont{Y.}~\bibnamefont{Soreq}}, \bibnamefont{and}
  \bibinfo{author}{\bibfnamefont{M.}~\bibnamefont{Spannowsky}},
  \bibinfo{journal}{JHEP} \textbf{\bibinfo{volume}{05}}, \bibinfo{pages}{145}
  (\bibinfo{year}{2015}), \eprint{1410.5440}.

\bibitem[{\citenamefont{Logan}(2015)}]{Logan:2014ppa}
\bibinfo{author}{\bibfnamefont{H.~E.} \bibnamefont{Logan}},
  \bibinfo{journal}{Phys. Rev.} \textbf{\bibinfo{volume}{D92}},
  \bibinfo{pages}{075038} (\bibinfo{year}{2015}), \eprint{1412.7577}.

\bibitem[{\citenamefont{Barger et~al.}(2012)\citenamefont{Barger, Ishida, and
  Keung}}]{Barger:2012hv}
\bibinfo{author}{\bibfnamefont{V.}~\bibnamefont{Barger}},
  \bibinfo{author}{\bibfnamefont{M.}~\bibnamefont{Ishida}}, \bibnamefont{and}
  \bibinfo{author}{\bibfnamefont{W.-Y.} \bibnamefont{Keung}},
  \bibinfo{journal}{Phys. Rev. Lett.} \textbf{\bibinfo{volume}{108}},
  \bibinfo{pages}{261801} (\bibinfo{year}{2012}), \eprint{1203.3456}.

\bibitem[{\citenamefont{Cao et~al.}(2017)\citenamefont{Cao, Chen, and
  Liu}}]{Cao:2016wib}
\bibinfo{author}{\bibfnamefont{Q.-H.} \bibnamefont{Cao}},
  \bibinfo{author}{\bibfnamefont{S.-L.} \bibnamefont{Chen}}, \bibnamefont{and}
  \bibinfo{author}{\bibfnamefont{Y.}~\bibnamefont{Liu}},
  \bibinfo{journal}{Phys. Rev.} \textbf{\bibinfo{volume}{D95}},
  \bibinfo{pages}{053004} (\bibinfo{year}{2017}), \eprint{1602.01934}.

\bibitem[{\citenamefont{Aaboud et~al.}(2018)}]{Aaboud:2018puo}
\bibinfo{author}{\bibfnamefont{M.}~\bibnamefont{Aaboud}} \bibnamefont{et~al.}
  (\bibinfo{collaboration}{ATLAS}), \bibinfo{journal}{Phys. Lett.}
  \textbf{\bibinfo{volume}{B786}}, \bibinfo{pages}{223} (\bibinfo{year}{2018}),
  \eprint{1808.01191}.

\bibitem[{\citenamefont{Sirunyan et~al.}(2019)}]{Sirunyan:2019twz}
\bibinfo{author}{\bibfnamefont{A.~M.} \bibnamefont{Sirunyan}}
  \bibnamefont{et~al.} (\bibinfo{collaboration}{CMS}), \bibinfo{journal}{Phys.
  Rev.} \textbf{\bibinfo{volume}{D99}}, \bibinfo{pages}{112003}
  (\bibinfo{year}{2019}), \eprint{1901.00174}.

\bibitem[{\citenamefont{Gunion and Haber}(1993)}]{Gunion:1992ce}
\bibinfo{author}{\bibfnamefont{J.~F.} \bibnamefont{Gunion}} \bibnamefont{and}
  \bibinfo{author}{\bibfnamefont{H.~E.} \bibnamefont{Haber}},
  \bibinfo{journal}{Phys. Rev.} \textbf{\bibinfo{volume}{D48}},
  \bibinfo{pages}{5109} (\bibinfo{year}{1993}).

\bibitem[{\citenamefont{Barger et~al.}(1997)\citenamefont{Barger, Berger,
  Gunion, and Han}}]{Barger:1996jm}
\bibinfo{author}{\bibfnamefont{V.~D.} \bibnamefont{Barger}},
  \bibinfo{author}{\bibfnamefont{M.~S.} \bibnamefont{Berger}},
  \bibinfo{author}{\bibfnamefont{J.~F.} \bibnamefont{Gunion}},
  \bibnamefont{and} \bibinfo{author}{\bibfnamefont{T.}~\bibnamefont{Han}},
  \bibinfo{journal}{Phys. Rept.} \textbf{\bibinfo{volume}{286}},
  \bibinfo{pages}{1} (\bibinfo{year}{1997}), \eprint{hep-ph/9602415}.

\bibitem[{\citenamefont{de~Florian et~al.}(2016)}]{deFlorian:2016spz}
\bibinfo{author}{\bibfnamefont{D.}~\bibnamefont{de~Florian}}
  \bibnamefont{et~al.} (\bibinfo{collaboration}{LHC Higgs Cross Section Working
  Group}) (\bibinfo{year}{2016}), \eprint{1610.07922}.

\bibitem[{\citenamefont{Brivio et~al.}(2019)\citenamefont{Brivio, Corbett, and
  Trott}}]{Brivio:2019myy}
\bibinfo{author}{\bibfnamefont{I.}~\bibnamefont{Brivio}},
  \bibinfo{author}{\bibfnamefont{T.}~\bibnamefont{Corbett}}, \bibnamefont{and}
  \bibinfo{author}{\bibfnamefont{M.}~\bibnamefont{Trott}}
  (\bibinfo{year}{2019}), \eprint{1906.06949}.

\bibitem[{\citenamefont{Georgi and Machacek}(1985)}]{Georgi:1985nv}
\bibinfo{author}{\bibfnamefont{H.}~\bibnamefont{Georgi}} \bibnamefont{and}
  \bibinfo{author}{\bibfnamefont{M.}~\bibnamefont{Machacek}},
  \bibinfo{journal}{Nucl. Phys.} \textbf{\bibinfo{volume}{B262}},
  \bibinfo{pages}{463} (\bibinfo{year}{1985}).

\bibitem[{\citenamefont{Chanowitz and Golden}(1985)}]{Chanowitz:1985ug}
\bibinfo{author}{\bibfnamefont{M.~S.} \bibnamefont{Chanowitz}}
  \bibnamefont{and} \bibinfo{author}{\bibfnamefont{M.}~\bibnamefont{Golden}},
  \bibinfo{journal}{Phys. Lett.} \textbf{\bibinfo{volume}{165B}},
  \bibinfo{pages}{105} (\bibinfo{year}{1985}).

\bibitem[{\citenamefont{Branco et~al.}(2012)\citenamefont{Branco, Ferreira,
  Lavoura, Rebelo, Sher, and Silva}}]{Branco:2011iw}
\bibinfo{author}{\bibfnamefont{G.~C.} \bibnamefont{Branco}},
  \bibinfo{author}{\bibfnamefont{P.~M.} \bibnamefont{Ferreira}},
  \bibinfo{author}{\bibfnamefont{L.}~\bibnamefont{Lavoura}},
  \bibinfo{author}{\bibfnamefont{M.~N.} \bibnamefont{Rebelo}},
  \bibinfo{author}{\bibfnamefont{M.}~\bibnamefont{Sher}}, \bibnamefont{and}
  \bibinfo{author}{\bibfnamefont{J.~P.} \bibnamefont{Silva}},
  \bibinfo{journal}{Phys. Rept.} \textbf{\bibinfo{volume}{516}},
  \bibinfo{pages}{1} (\bibinfo{year}{2012}), \eprint{1106.0034}.

\bibitem[{\citenamefont{Profumo et~al.}(2007)\citenamefont{Profumo,
  Ramsey-Musolf, and Shaughnessy}}]{Profumo:2007wc}
\bibinfo{author}{\bibfnamefont{S.}~\bibnamefont{Profumo}},
  \bibinfo{author}{\bibfnamefont{M.~J.} \bibnamefont{Ramsey-Musolf}},
  \bibnamefont{and}
  \bibinfo{author}{\bibfnamefont{G.}~\bibnamefont{Shaughnessy}},
  \bibinfo{journal}{JHEP} \textbf{\bibinfo{volume}{08}}, \bibinfo{pages}{010}
  (\bibinfo{year}{2007}), \eprint{0705.2425}.

\bibitem[{\citenamefont{Agashe et~al.}(2005)\citenamefont{Agashe, Contino, and
  Pomarol}}]{Agashe:2004rs}
\bibinfo{author}{\bibfnamefont{K.}~\bibnamefont{Agashe}},
  \bibinfo{author}{\bibfnamefont{R.}~\bibnamefont{Contino}}, \bibnamefont{and}
  \bibinfo{author}{\bibfnamefont{A.}~\bibnamefont{Pomarol}},
  \bibinfo{journal}{Nucl. Phys.} \textbf{\bibinfo{volume}{B719}},
  \bibinfo{pages}{165} (\bibinfo{year}{2005}), \eprint{hep-ph/0412089}.

\bibitem[{\citenamefont{Chacko et~al.}(2006)\citenamefont{Chacko, Goh, and
  Harnik}}]{Chacko:2005pe}
\bibinfo{author}{\bibfnamefont{Z.}~\bibnamefont{Chacko}},
  \bibinfo{author}{\bibfnamefont{H.-S.} \bibnamefont{Goh}}, \bibnamefont{and}
  \bibinfo{author}{\bibfnamefont{R.}~\bibnamefont{Harnik}},
  \bibinfo{journal}{Phys. Rev. Lett.} \textbf{\bibinfo{volume}{96}},
  \bibinfo{pages}{231802} (\bibinfo{year}{2006}), \eprint{hep-ph/0506256}.

\bibitem[{\citenamefont{Craig et~al.}(2015{\natexlab{a}})\citenamefont{Craig,
  Katz, Strassler, and Sundrum}}]{Craig:2015pha}
\bibinfo{author}{\bibfnamefont{N.}~\bibnamefont{Craig}},
  \bibinfo{author}{\bibfnamefont{A.}~\bibnamefont{Katz}},
  \bibinfo{author}{\bibfnamefont{M.}~\bibnamefont{Strassler}},
  \bibnamefont{and} \bibinfo{author}{\bibfnamefont{R.}~\bibnamefont{Sundrum}},
  \bibinfo{journal}{JHEP} \textbf{\bibinfo{volume}{07}}, \bibinfo{pages}{105}
  (\bibinfo{year}{2015}{\natexlab{a}}), \eprint{1501.05310}.

\bibitem[{\citenamefont{O'Connell et~al.}(2007)\citenamefont{O'Connell,
  Ramsey-Musolf, and Wise}}]{OConnell:2006rsp}
\bibinfo{author}{\bibfnamefont{D.}~\bibnamefont{O'Connell}},
  \bibinfo{author}{\bibfnamefont{M.~J.} \bibnamefont{Ramsey-Musolf}},
  \bibnamefont{and} \bibinfo{author}{\bibfnamefont{M.~B.} \bibnamefont{Wise}},
  \bibinfo{journal}{Phys. Rev.} \textbf{\bibinfo{volume}{D75}},
  \bibinfo{pages}{037701} (\bibinfo{year}{2007}), \eprint{hep-ph/0611014}.

\bibitem[{\citenamefont{Ramsey-Musolf et~al.}(2021)\citenamefont{Ramsey-Musolf,
  Yu, and Zhou}}]{Ramsey-Musolf:2021ldh}
\bibinfo{author}{\bibfnamefont{M.~J.} \bibnamefont{Ramsey-Musolf}},
  \bibinfo{author}{\bibfnamefont{J.-H.} \bibnamefont{Yu}}, \bibnamefont{and}
  \bibinfo{author}{\bibfnamefont{J.}~\bibnamefont{Zhou}}
  (\bibinfo{year}{2021}), \eprint{2104.10709}.

\bibitem[{\citenamefont{Aad et~al.}(2019)}]{Aad:2019mbh}
\bibinfo{author}{\bibfnamefont{G.}~\bibnamefont{Aad}} \bibnamefont{et~al.}
  (\bibinfo{collaboration}{ATLAS}) (\bibinfo{year}{2019}), \eprint{1909.02845}.

\bibitem[{\citenamefont{Kling et~al.}(2016)\citenamefont{Kling, No, and
  Su}}]{Kling:2016opi}
\bibinfo{author}{\bibfnamefont{F.}~\bibnamefont{Kling}},
  \bibinfo{author}{\bibfnamefont{J.~M.} \bibnamefont{No}}, \bibnamefont{and}
  \bibinfo{author}{\bibfnamefont{S.}~\bibnamefont{Su}}, \bibinfo{journal}{JHEP}
  \textbf{\bibinfo{volume}{09}}, \bibinfo{pages}{093} (\bibinfo{year}{2016}),
  \eprint{1604.01406}.

\bibitem[{\citenamefont{Eriksson et~al.}(2010)\citenamefont{Eriksson, Rathsman,
  and Stal}}]{Eriksson:2009ws}
\bibinfo{author}{\bibfnamefont{D.}~\bibnamefont{Eriksson}},
  \bibinfo{author}{\bibfnamefont{J.}~\bibnamefont{Rathsman}}, \bibnamefont{and}
  \bibinfo{author}{\bibfnamefont{O.}~\bibnamefont{Stal}},
  \bibinfo{journal}{Comput. Phys. Commun.} \textbf{\bibinfo{volume}{181}},
  \bibinfo{pages}{189} (\bibinfo{year}{2010}), \eprint{0902.0851}.

\bibitem[{\citenamefont{Chowdhury and Eberhardt}(2018)}]{Chowdhury:2017aav}
\bibinfo{author}{\bibfnamefont{D.}~\bibnamefont{Chowdhury}} \bibnamefont{and}
  \bibinfo{author}{\bibfnamefont{O.}~\bibnamefont{Eberhardt}},
  \bibinfo{journal}{JHEP} \textbf{\bibinfo{volume}{05}}, \bibinfo{pages}{161}
  (\bibinfo{year}{2018}), \eprint{1711.02095}.

\bibitem[{\citenamefont{Bélusca-Maïto
  et~al.}(2017)\citenamefont{Bélusca-Maïto, Falkowski, Fontes, Romão, and
  Silva}}]{Belusca-Maito:2016dqe}
\bibinfo{author}{\bibfnamefont{H.}~\bibnamefont{Bélusca-Maïto}},
  \bibinfo{author}{\bibfnamefont{A.}~\bibnamefont{Falkowski}},
  \bibinfo{author}{\bibfnamefont{D.}~\bibnamefont{Fontes}},
  \bibinfo{author}{\bibfnamefont{J.~C.} \bibnamefont{Romão}},
  \bibnamefont{and} \bibinfo{author}{\bibfnamefont{J.~P.} \bibnamefont{Silva}},
  \bibinfo{journal}{Eur. Phys. J.} \textbf{\bibinfo{volume}{C77}},
  \bibinfo{pages}{176} (\bibinfo{year}{2017}), \eprint{1611.01112}.

\bibitem[{\citenamefont{Hartling et~al.}(2014)\citenamefont{Hartling, Kumar,
  and Logan}}]{Hartling:2014xma}
\bibinfo{author}{\bibfnamefont{K.}~\bibnamefont{Hartling}},
  \bibinfo{author}{\bibfnamefont{K.}~\bibnamefont{Kumar}}, \bibnamefont{and}
  \bibinfo{author}{\bibfnamefont{H.~E.} \bibnamefont{Logan}}
  (\bibinfo{year}{2014}), \eprint{1412.7387}.

\bibitem[{\citenamefont{Li et~al.}(2019)\citenamefont{Li, Xu, Yu, and
  Zhu}}]{Li:2019ghf}
\bibinfo{author}{\bibfnamefont{H.-L.} \bibnamefont{Li}},
  \bibinfo{author}{\bibfnamefont{L.-X.} \bibnamefont{Xu}},
  \bibinfo{author}{\bibfnamefont{J.-H.} \bibnamefont{Yu}}, \bibnamefont{and}
  \bibinfo{author}{\bibfnamefont{S.-H.} \bibnamefont{Zhu}},
  \bibinfo{journal}{JHEP} \textbf{\bibinfo{volume}{09}}, \bibinfo{pages}{010}
  (\bibinfo{year}{2019}), \eprint{1904.05359}.

\bibitem[{\citenamefont{Alonso et~al.}(2016)\citenamefont{Alonso, Jenkins, and
  Manohar}}]{Alonso:2016btr}
\bibinfo{author}{\bibfnamefont{R.}~\bibnamefont{Alonso}},
  \bibinfo{author}{\bibfnamefont{E.~E.} \bibnamefont{Jenkins}},
  \bibnamefont{and} \bibinfo{author}{\bibfnamefont{A.~V.}
  \bibnamefont{Manohar}}, \bibinfo{journal}{Phys. Lett.}
  \textbf{\bibinfo{volume}{B756}}, \bibinfo{pages}{358} (\bibinfo{year}{2016}),
  \eprint{1602.00706}.

\bibitem[{\citenamefont{Bellazzini et~al.}(2014)\citenamefont{Bellazzini,
  Csaki, and Serra}}]{Bellazzini:2014yua}
\bibinfo{author}{\bibfnamefont{B.}~\bibnamefont{Bellazzini}},
  \bibinfo{author}{\bibfnamefont{C.}~\bibnamefont{Csaki}}, \bibnamefont{and}
  \bibinfo{author}{\bibfnamefont{J.}~\bibnamefont{Serra}},
  \bibinfo{journal}{Eur. Phys. J.} \textbf{\bibinfo{volume}{C74}},
  \bibinfo{pages}{2766} (\bibinfo{year}{2014}), \eprint{1401.2457}.

\bibitem[{\citenamefont{Liu et~al.}(2018)\citenamefont{Liu, Low, and
  Yin}}]{Liu:2018vel}
\bibinfo{author}{\bibfnamefont{D.}~\bibnamefont{Liu}},
  \bibinfo{author}{\bibfnamefont{I.}~\bibnamefont{Low}}, \bibnamefont{and}
  \bibinfo{author}{\bibfnamefont{Z.}~\bibnamefont{Yin}},
  \bibinfo{journal}{Phys. Rev. Lett.} \textbf{\bibinfo{volume}{121}},
  \bibinfo{pages}{261802} (\bibinfo{year}{2018}), \eprint{1805.00489}.

\bibitem[{\citenamefont{Liu et~al.}(2017)\citenamefont{Liu, Low, and
  Wagner}}]{Liu:2017dsz}
\bibinfo{author}{\bibfnamefont{D.}~\bibnamefont{Liu}},
  \bibinfo{author}{\bibfnamefont{I.}~\bibnamefont{Low}}, \bibnamefont{and}
  \bibinfo{author}{\bibfnamefont{C.~E.~M.} \bibnamefont{Wagner}},
  \bibinfo{journal}{Phys. Rev.} \textbf{\bibinfo{volume}{D96}},
  \bibinfo{pages}{035013} (\bibinfo{year}{2017}), \eprint{1703.07791}.

\bibitem[{\citenamefont{Craig et~al.}(2015{\natexlab{b}})\citenamefont{Craig,
  Knapen, and Longhi}}]{Craig:2014aea}
\bibinfo{author}{\bibfnamefont{N.}~\bibnamefont{Craig}},
  \bibinfo{author}{\bibfnamefont{S.}~\bibnamefont{Knapen}}, \bibnamefont{and}
  \bibinfo{author}{\bibfnamefont{P.}~\bibnamefont{Longhi}},
  \bibinfo{journal}{Phys. Rev. Lett.} \textbf{\bibinfo{volume}{114}},
  \bibinfo{pages}{061803} (\bibinfo{year}{2015}{\natexlab{b}}),
  \eprint{1410.6808}.

\bibitem[{\citenamefont{Xu et~al.}(2020)\citenamefont{Xu, Yu, and
  Zhu}}]{Xu:2018ofw}
\bibinfo{author}{\bibfnamefont{L.-X.} \bibnamefont{Xu}},
  \bibinfo{author}{\bibfnamefont{J.-H.} \bibnamefont{Yu}}, \bibnamefont{and}
  \bibinfo{author}{\bibfnamefont{S.-H.} \bibnamefont{Zhu}},
  \bibinfo{journal}{Phys. Rev. D} \textbf{\bibinfo{volume}{101}},
  \bibinfo{pages}{095014} (\bibinfo{year}{2020}), \eprint{1810.01882}.

\end{thebibliography}

\end{document}